\documentclass[a4paper,11pt]{article}
\pdfoutput=1
\usepackage{jheppub,booktabs}
\usepackage{amsmath,amssymb}
\usepackage{cancel}
\usepackage{mathrsfs}
\usepackage{graphicx}
\usepackage{hepunits}
\usepackage{wasysym}
\usepackage{feynmf}
\usepackage[utf8x]{inputenc}

\author{Adrián Carmona}
\author{and Florian Goertz}
\affiliation{
Institute for Theoretical Physics, \\
ETH Zurich, 8093 Zurich, Switzerland}
\emailAdd{carmona@itp.phys.ethz.ch}
\emailAdd{fgoertz@itp.phys.ethz.ch}
\title{Custodial Leptons and Higgs Decays}
\abstract{
We study the effects of extended fermion sectors, respecting custodial symmetry, on Higgs production and decay. The resulting protection for the $Z \to b_L b_L$ and $Z \to \tau_R \tau_R$ decays allows for potentially interesting signals in Higgs physics, while maintaining the good agreement of the Standard Model with precision tests, without significant fine-tuning.
Although being viable setups on their own, the models we study can particularly be motivated as the low energy effective theories of the composite Higgs models MCHM$_{5}$ and MCHM$_{10}$ or the corresponding gauge-Higgs unification models. The spectra can be identified with the light custodians present in these theories. These have the potential to describe the relevant physics in their fermion sectors in a simplified and transparent way. In contrast to previous studies of composite models, we consider the impact of a realistic lepton sector on the Higgs decays. We find significant modifications in the decays to $\tau$ leptons and photons due to the new leptonic resonances. While from a pure low energy perspective an enhancement of the channel $pp \to h \to \gamma\gamma$ turns out to be possible, if one considers constraints on the parameters from the full structure of the composite models, the decay mode into photons is always reduced. We also demonstrate that taking into account the non-linearity of the Higgs sector does not change the qualitative picture for the decays into $\tau$ leptons or photons in the case of the dominant Higgs production mechanism.
}
\date{\today}
\begin{document}

\maketitle
\section{Introduction}

Recently, a new boson has been discovered in both the ATLAS \cite{aad:2012gk} and CMS \cite{serguei:2012gu} experiments at the Large Hadron Collider (LHC). It is still to be confirmed that this particle is the long-sought Higgs boson, the last missing ingredient of the Standard Model (SM) of Particle Physics. While the overall picture of the measured cross sections in its various decay channels is in reasonable agreement with the SM-Higgs expectations, there is also still room for significant deviations. In particular,
 both experiments observe a tendency towards an enhanced decay into two photons, at the level of 1-2 $\sigma$. Moreover, although the trend for a depletion in the decay-channel into $\tau$ leptons has become less strong, a reduced rate in that mode still fits well with the data - the ATLAS results are still compatible with a vanishing signal at the level of about $1\,\sigma$ \cite{ATLASnote,CMSnote}.

These trends might vanish after more statistics has been accumulated, however it is always worth studying the impact of extensions of the SM on Higgs physics, to examine to what extend they could agree with experimental tendencies or to constrain these models. The most straightforward way to enhance for example the two-photon signal without affecting other channels too much is to add new leptons to the SM with Higgs couplings that are not aligned to their masses. This opens the possibility of a constructive interference of the new leptons in the loop contributing to $h\to \gamma\gamma$ with the $W^\pm$ boson loop. This option has been considered in \cite{Joglekar:2012vc,ArkaniHamed:2012kq}.
In general, many models involving new particles have been introduced to account for the enhancement in the photon channel, some without embedding the new physics into a motivated UV completion,
others studying $h\to\gamma\gamma$ in (more) complete models, addressing the hierarchy problem of particle physics \cite{Ellwanger:2011aa, Goertz:2011hj, Blum:2012kn, Cao:2012fz, Boudjema:2012cq, Wang:2012zv, Bellazzini:2012tv, Dawson:2012di, Azatov:2012wq, Bonne:2012im, Hagiwara:2012mga, Bellazzini:2012mh, Benbrik:2012rm, Buckley:2012em, An:2012vp, Alves:2012yp, Abe:2012fb, Bertolini:2012gu, Joglekar:2012vc, ArkaniHamed:2012kq, Craig:2012vn, Almeida:2012bq, Batell:2012mj, Hashimoto:2012qe, SchmidtHoberg:2012yy, Reece:2012gi, Boudjema:2012in,Wang:2012ts, Alves:2012fb, Bertuzzo:2012bt, Chala:2012af, Dawson:2012mk, Haisch:2012re, Bechtle:2012jw, Petersson:2012nv, SchmidtHoberg:2012ip, Dudas:2012fa, Berg:2012cg, Iltan:2012pr, Chao:2012xt, Han:2012dd, Chun:2013ft, Funatsu:2013ni, Grojean:2013kd, Fan:2013qn}.

In this paper we want to consider, on the one hand, simple low-energy models featuring new quarks and leptons, that allow to clearly keep track of the observed effects in Higgs physics. On the other hand we also try to address the question where the new particles could come from, thereby increasing the predictivity of the setup. Most importantly, we are led by the request to keep the good agreement of the SM with precision tests also for the extended setup, without introducing severe fine-tuning. To that extend, we embed the new physics sector in a way that respects a custodial symmetry protecting the $T$ parameter as well as the couplings of the $Z$ boson to fermions. Such a symmetry is likely to be an ingredient of viable new-physics models at the TeV scale.
 
To be specific, we study two realizations of an extended fermion sector, one featuring fundamental representations of $SO(5)$, the other employing also an adjoint (ten-dimensional) representation of the same Lie-group, which both possess a custodial $SU(2)_L \times SU(2)_R \times P_{LR}\ (\subset SO(5)$) symmetry. Although being valid setups on their own, they are particularly motivated as the low energy tails of minimal composite Higgs models (MCHM) or corresponding models of gauge-Higgs unification (GHU) \cite{Manton:1979kb, Hatanaka:1998yp, vonGersdorff:2002as, Csaki:2002ur, Contino:2003ve, Agashe:2004rs}.

Such models lead generically to the presence of light resonances associated to the top quark and required by custodial symmetry~\cite{Carena:2006bn, Carena:2007ua, Contino:2006qr, Carena:2007tn, Pomarol:2008bh, Panico:2010is, Matsedonskyi:2012ym}, with masses significantly below the actual scale of these models, $m_{\rm cust} \ll f$. This is a consequence of the large value of the top mass and the enlarged fermion representations chosen to protect the $Zb\bar{b}$ vertex from anomalous corrections.  Looking at the value of the lepton masses, there is \emph{a priori} no reason to think that something analogous could happen in the lepton sector. However, as it was shown in \cite{delAguila:2010vg}, trying to explain the observed pattern of lepton masses and mixings with the help of a discrete $A_4$ symmetry requires the $\tau$ to be more composite than naively expected and thus makes the appearance of light $\tau$ custodians quite likely. Phenomenological consequences of such resonances at the LHC were also studied in detail in \cite{delAguila:2010es}. 
Although the presence of $\tau$ custodians was predicted in \cite{delAguila:2010vg} just for the MCHM$_5$, where all the leptons are in fundamental representations of $SO(5)\times U(1)_X$, they are also present when we choose larger representations for the charged leptons \cite{next}. Therefore, finding light $\tau$ custodians at the LHC, directly or through the modifications induced on the different Higgs decays, could be interpreted as a strong hint for the compositeness of the recently discovered Higgs boson. 

At low energies $E\ll f$ one would see the SM plus resonances coming with the top quark and the $\tau$ lepton, which just corresponds to what we will be studying in this article. The impact of possible UV completions on the parameters of the models will be detailed further below. We will use the abbreviations MCHM$_5$ (MCHM$_{5+10}$) for these extended fermion sectors featuring \textbf{5}s (\textbf{5}s plus one \textbf{10}) of $SO(5)$, although we will not consider the full composite models, {\it i.e.} we will neglect heavier fermionic resonances, possible changes in the gauge-boson sector and for the first part of the analysis also the effects from the non-linearity of the Higgs sector. We will however study the impact of the latter effect at the end of the article. Note that, in what we call the MCHM$_{5+10}$, we will only embed the $\tau_R$ in a \textbf{10}, whereas the other fermions will remain in the fundamental representation of $SO(5)$. 
Due to the significant compositeness of the $\tau$ lepton one expects non-negligible effects in the lepton sector of Higgs physics, which have been neglected so far. It would be interesting to consider these effects also in complete composite Higgs models. A detailed examination of this is left for future work \cite{next}.  However, as we will explain below, the simplified setup of our analysis will already grasp the most important structure of these effects in a very transparent way and moreover is also valid in a more general context.

In Section \ref{sec:LE}, we will detail the extended fermion sectors studied in this work and derive the corresponding spectra and Higgs couplings. The anatomy of the Higgs-production and decay cross sections in the models at hand will be studied in Section \ref{sec:Higgs}, where we also give numerical predictions for various search channels. Finally, our conclusions will be presented in Section~\ref{sec:conclusions}.

\section{Low Energy Spectrum and Higgs Couplings of the Models}
\label{sec:LE}

The emergence of light leptonic custodians in the MCHM$_5$ has been motivated in \cite{delAguila:2010vg} from a UV perspective for a complete composite Higgs model, and similar considerations hold, putting the $\tau_R$ into a \textbf{10} of $SO(5)$ \cite{next}. 
However, as mentioned before, the setup for our analysis of Higgs production and decay will only be the corresponding low energy theory, including the light custodians.
Beyond that, we do not even have to rely on a certain UV completion of this model but rather consider it as a general low energy setup featuring a viable implementation of custodial protection, with the only additional assumption being that the scale of a possible UV completion is significantly larger than the mass-scale of the new fermionic resonances considered. The particle spectrum is then inspired by the prominent role of the third fermion generation.
This will be the starting point for our analysis. 

In the case we do want to consider a model of GHU (or composite Higgs) completing this setup and causing the existence of the light resonances, this has however to be taken with a grain of salt. The 5D structure of this model leads to relations between different Kaluza-Klein (KK) modes of the same level such that the suppressed contributions of modes with masses significantly larger than the light custodians can be lifted by relatively larger Yukawa couplings in the triangle diagrams examined in Section \ref{sec:Higgs}. To consider really the full structure of the fermion sectors of these models at leading order, it seems important to take into account complete KK levels.\footnote{Note that in full 5D models all leptons live in representations of $SO(5)$.} In the models that we will study, this is indeed assured (effectively). In one case, those heavy resonances, that are missing to complete the KK level of the light custodians that we consider, will have negligible couplings to the Higgs boson. In the other, it turns out that already those modes present in our low energy setup exhibit the structure that describes the full KK structure. The former is true for the $Y$ sector in the MCHM$_{5+10}$, while the latter happens for the top, bottom and $\tau$ sectors. We will elaborate more on this further below. We now leave again the question of a possible UV completion for the next considerations. 

In chosing how to embed the SM fermion sector in enlarged representations, we are led by the assumption that the (right handed) top quark plays a special role in the fermion sector and that thus the new light resonances should complete the right handed top to form a representation under $SO(5)$ and that the same could hold for the right handed $\tau$ sector. The assumption that (only) those two SM fermions couple strongly to the new sector will also constrain the ranges that we will chose for the parameters of the models, see below. 

In consequence, the setup we want to consider for the following analysis corresponds to the SM Lagrangian, supplemented only with vector-like fermions, associated to the top and $\tau$ sectors, that live in fundamental or adjoint representations of $SO(5)$, such that there is a custodial protection for $Z\to b_L b_L$ as well as $Z \to \tau_R \tau_R$ decays (which is important due to the non-negligible mixings with the new sector). 
We will now give the details of this setup, with the focus on the spectrum and the Higgs couplings. We start with the option of putting the fermions in the fundamental representation of $SO(5)$. As mentioned, we will call this setup MCHM$_5$ in the following, although for us it is only the low energy theory, featuring fermions in the fundamental representation of $SO(5)$ and not a complete composite model. For the lepton sector, this model has been studied in \cite{delAguila:2010es} and we will give a short review of the key features in the following, generalizing the setup to include quarks \cite{Atre:2008iu,Atre:2011ae}. After that we will spell out the low energy theory corresponding to the option of putting fermions in the adjoint representation, a \textbf{10} of $SO(5)$ (MCHM$_{10}$).

\subsection{MCHM$_5$}
\label{sec:MCHM5}

The light $\tau$ custodians present in this model in addition to the SM fermions are contained in the lepton-multiplets \cite{delAguila:2010es}
\begin{equation}
\label{eq:2Doub}
L_{1L,R}^{(0)}=\begin{pmatrix}N_{1L,R}^{(0)}\\ E_{1L,R}^{(0)}\end{pmatrix}\sim \mathbf{2}_{-\frac{1}{2}},\quad 
L_{2L,R}^{(0)}=\begin{pmatrix} E_{2L,R}^{(0)}
\\ Y_{2L,R}^{(0)} \end{pmatrix}\sim \mathbf{2}_{-\frac{3}{2}}\,,
\end{equation}
where the given transformation properties correspond to ${(SU(2)_L)}_Y$ and the $SU(2)_R$ quantum numbers are $T_R^3=1/2$ and $T_R^3=-1/2$, respectively, following from the embedding in the full $SO(5)\times U(1)_X$ gauge group.  The superscript $(0)$ indicates the current basis. The model is designed such that the custodial symmetry protects the $Z \tau_R \tau_R$ coupling (see \cite{Agashe:2006at}). In addition, we assume a similar embedding of the quark sector, now featuring a protection for the $Z b_L b_L$ coupling. This is achieved by a setup which, due to the large top mass, leads to the light custodians
\begin{equation}
\label{eq:2Doubq}
Q_{1L,R}^{(0)}=\begin{pmatrix}\Lambda_{1L,R}^{(0)}\\ T_{1L,R}^{(0)}\end{pmatrix}\sim \mathbf{2}_{\frac{7}{6}},\quad 
Q_{2L,R}^{(0)}=\begin{pmatrix} T_{2L,R}^{(0)}
\\ B_{2L,R}^{(0)} \end{pmatrix}\sim \mathbf{2}_{\frac{1}{6}}\,.
\end{equation}

The Lagrangian of our model consists of the SM operators, supplemented with all possible gauge invariant combinations involving the new fermion multiplets.
Neglecting the first two generations, which are assumed to have negligible couplings to the new resonances, the mass and Yukawa couplings are given by 
\begin{equation}
\label{eq:Ll}
{\cal L}_{L} =
- y_l\, \bar{l}_L^{(0)}\varphi \tau^{(0)}_R
-y_l^\prime
\Big[\bar{L}^{(0)}_{1L} \varphi + \bar{L}^{(0)}_{2L}
  \tilde{\varphi} \Big] \tau^{(0)}_R 
- M_l \Big[ \bar{L}^{(0)}_{1L} L^{(0)}_{1R}+ \bar{L}^{(0)}_{2L}
  L^{(0)}_{2R} \Big]+\mathrm{h.c.}
\end{equation}
and
\begin{equation}
\label{eq:Lq}
{\cal L}_{Q} =
- y_q\, \bar{q}_L^{(0)}\varphi t^{(0)}_R
- y_q^\prime
\Big[\bar{Q}^{(0)}_{1L} \varphi + \bar{Q}^{(0)}_{2L}
  \tilde{\varphi} \Big] t^{(0)}_R 
- M_Q \Big[ \bar{Q}^{(0)}_{1L} Q^{(0)}_{1R}+ \bar{Q}^{(0)}_{2L}
  Q^{(0)}_{2R} \Big]+\mathrm{h.c.}\,,
\end{equation}
where $l_L^{(0)}$ and $\tau_R^{(0)}$ ($q_L^{(0)}$ and $t_R^{(0)}$) denote the third generation SM leptons (quarks). After electroweak symmetry breaking (EWSB) and in unitary gauge, the Higgs doublet is given by $\varphi=1/\sqrt2\, (0,v+h)^T$, with $v=246$\,GeV and $h$ the Higgs boson, whereas $\tilde \varphi = i \sigma_2 \varphi^\ast$.
Note that we neglected the couplings of the right handed bottom quark (or the corresponding neutrino), which are SM like since there are no new resonances to which it could couple, due to the charges and multiplet structure of the MCHM$_5$. The fact that different operators above have the same Yukawa couplings or vector-like masses is due to the $P_{LR}$ symmetry, exchanging $SU(2)_L \leftrightarrow SU(2)_R$.

The model is simple enough that compact analytical formulas can be derived for the physical masses and Higgs couplings of the extended fermion sector, which we will give here for the leptons. For the quarks, the same formulas hold with the replacements
$\tau \to t, E \to T, Y \to B,$ and $N\to \Lambda$ and we will suppress indices if convenient.
The only non-trivial fermion mass matrix (featuring non-vanishing Yukawa couplings) that follows from (\ref{eq:Ll}) belongs to the E (T) sector and reads
\begin{equation}
\label{eq:Mm}
\mathcal{M}^5=\frac{v}{\sqrt2}\begin{pmatrix}
y & 0 & 0 \\ y^\prime &  \frac{\sqrt 2}{v} M & 0 \\ y^\prime & 0 & \frac{\sqrt 2}{v} M  \end{pmatrix}\,.
\end{equation}
Let us already make a first comment on the expected size of the entries of this matrix. 
Our working assumption is that the top-quark and the $\tau$ lepton couple significantly, with a strength governed by the electroweak scale $v$, to the new physics and thus we expect $y^\prime \sim 1$. The parameter $y$ describes the mass term between the SM-like top (or $\tau$) fields and is thus governed to a large extend by their mass eigenvalues. The vector-like masses of the new resonances associated to the top and $\tau$ sectors are, motivated by the exposed role of these fermions, expected to be significantly smaller than the general scale of new physics, {\it i.e.}, $\mathcal{O}({\rm TeV})\gg M \gg v$.

From the explicit perspective of a composite model for example, one expects the vector-like mass $M$ to be lighter than the scale of compositeness $f\gg M\gg v$. The Yukawa coupling $y$ parametrizes the interactions of the right handed $\tau$ and $t$, which have a sizable composite component, with the composite Higgs boson and their more elementary left-handed components, whereas $y^\prime$ describes interactions of the $\tau_R$ or $t_R$ with heavy composite resonances and the composite Higgs. In consequence both mass-couplings are non negligible, however one still expects typically $v\,y \ll v\,y^\prime \ll M$, where the first ``$\ll$'' should rather be a ``$<$''  for the top-quark sector. Remember that the flavor pattern of composite Higgs models (featuring partial compositeness) matches nicely with the experimental observation that possible deviations from the SM in the third generation of fermions are less severely constrained. Beyond these considerations, note that already from the pure fact that no additional fermions have been found at the LHC yet, one expects $M\gg v$ (see (\ref{eq:ME}) below).
We will use such insights on hierarchies in chosing the parameter-space for our scans in Section \ref{sec:pheno}.

The matrix ($\ref{eq:Mm}$) can be diagonalized via a bi-unitary transformation, 
$U_L^{5\dagger} \mathcal{M}^5 U_R^5 =
\mathcal{M}^5_{\mbox{diag}} = (m_\tau, m_{E_1},m_{E_2})$, 
which due to the structure of (\ref{eq:Mm}) takes the simple form
\begin{equation}
\label{eq:rot}
U_{L,R}^5=
\begin{pmatrix}
c_{L,R} & 0 & s_{L,R} \\
-\frac{s_{L,R}}{\sqrt{2}}
 & \frac{1}{\sqrt{2}} &  \frac{c_{L,R}}{\sqrt{2}} \\
-\frac{s_{L,R}}{\sqrt{2}}
 & -\frac{1}{\sqrt{2}} &  \frac{c_{L,R}}{\sqrt{2}} 
\end{pmatrix},
\end{equation}
with the sine and cosine of the mixing angles $s_{L,R} \equiv \sin (\theta_{L,R})$, $c_{L,R} \equiv \cos
(\theta_{L,R})$. 
The relevant input parameters of the model at this point are $y$, $y^\prime$
and $M$.  However, it will be more convenient to use as input
$m_\tau$, $s_R$, and $M$, where the first quantity is already fixed by experiment.
The left-handed mixing parameter is related to them via
\begin{equation}
s_L= s_R \frac{m_\tau}{M}.
\end{equation}

The  physical non-SM  states consist of three heavy particles of degenerate vector-like mass 
\begin{equation}
\label{eq:ME}
m_N=m_{E_1}=m_Y=M,
\end{equation}
and electric charges of $Q=0,-1,-2$ ($Q=5/3,2/3,-1/3$) in the lepton (quark) sector.
In addition, there is a heavier $Q=-1$ ($Q=2/3$) state with
\begin{equation}
m_{E_2}=\frac{M}{c_R}\sqrt{1-s_R^2\frac{m_\tau^2}{M^2}}.
\end{equation}

The couplings of the fermions to the Higgs boson are given by
\begin{equation}
\label{eq:gH5}
{\cal L}_h=\sum_{f=E,T}\bar \Psi_L^{f5\,(0)} g_{h5}^{f(0)} \Psi_R^{f5\,(0)}  h+ {\rm h.c.} \,,
\end{equation}
where $\Psi^{E5\,(0)}\equiv (\tau^{(0)},E_1^{(0)},E_2^{(0)})^T$, $\Psi^{T5\,(0)}\equiv (t^{(0)},T_1^{(0)},T_2^{(0)})^T$ and
\begin{equation}
\sqrt2\, g_{h5}^{f(0)}=\begin{pmatrix}
y & 0 & 0 \\ y^\prime & 0 & 0 \\ y^\prime & 0 & 0 \end{pmatrix}\,.
\end{equation}
After rotating to the diagonal mass basis, the Higgs-coupling matrix with leptons becomes
\begin{equation}
\label{eq:gh5}
g_{h5}^E=U_L^{5\dagger} g_{h5}^{E(0)} U_R^5=\frac 1 v
\begin{pmatrix}
c_R^2 m_\tau & 0 & s_R c_R m_\tau \\
0 & 0 & 0 \\
s_R c_R M_{E_2}
& 0 & 
s_R^2 M_{E_2}
\end{pmatrix}\,,
\end{equation}
and similarly for the quark sector.
It features off-diagonal entries, due to the presence of the vector-like masses $M$.
Note that in the MCHM$_{5}$ the new resonances belonging to the $N,Y,\Lambda,B$ sectors do not couple to the Higgs, as one can not write a gauge invariant term mediating such a coupling.

\subsection{MCHM$_{5+10}$}

It is easy to show that, analogously to the discussion in \cite{delAguila:2010vg}, embedding the $\tau_R$ in an adjoint representation of $SO(5)$ and requiring a custodial protection for the $Z \tau_R \tau_R$ coupling leads to the following light custodians \cite{next}, belonging to a \textbf{10} of $SO(5)$,\footnote{With some abuse of notation, we use the same names as already used for the MCHM$_5$.  However, the assignment will be clear from the context.}
\begin{eqnarray}
L_{1L,R}^{(0)}&=&\begin{pmatrix}N_{1L,R}^{(0)}\\ E_{1L,R}^{(0)}\end{pmatrix}\sim \mathbf{2}_{-\frac{1}{2}},\quad 
L_{2L,R}^{(0)}=\begin{pmatrix} E_{2L,R}^{(0)}
\\ Y_{2L,R}^{(0)} \end{pmatrix}\sim \mathbf{2}_{-\frac{3}{2}}, \\
L_{3L,R}^{(0)}&=&\begin{pmatrix} N_{3L,R}^{(0)}\\E_{3L,R}^{(0)}
\\ Y_{3L,R}^{(0)} \end{pmatrix}\sim \mathbf{3}_{-1}, \quad N_{2L,R}^{(0)}\sim \mathbf{1}_0,\quad Y_{1L,R}^{(0)}\sim \mathbf{1}_{-2}.
\end{eqnarray}
Note that we will keep the other lepton multiplets, as well as the quarks, in the fundamental representation.\footnote{For the low energy Lagrangian of the light custodians, given in this section, the embedding of the first two generations is irrelevant.} As a consequence, we do not give the quark sector for the \textbf{10}, which can however be worked out straightforwardly. This setup can be seen as the most straightforward departure from the MCHM$_5$ and, as we will detail further below, a first step towards the embedding of all SM fermions into \textbf{10}s of $SO(5)$ in a GHU model, while still describing the full fermion sector of the composite model by the light custodians in a simple and self-contained way.

Here, again, the superscript $(0)$ indicates the current basis and the $SU(2)_R$ quantum numbers are $T_R^3=1/2,-1/2$ for the two $SU(2)_L$ doublets,
whereas the $SU(2)_L$ triplets are $SU(2)_R$ singlets and vice versa. The relevant part of the Yukawa and mass Lagrangian now reads\footnote{Note that we will neglect the neutrino sector, as it is irrelevant for the following discussions.}
\begin{eqnarray}
\mathcal{L}&=&-y\, \bar{l}_L^{(0)}\varphi \tau_R^{(0)}
%-\sqrt 2 y_\nu\, \bar{l}_L^{(0)}\tilde\varphi \nu_R^{(0)}
- y^\prime \left[\bar{L}_{1L}^{(0)}\varphi+\bar{L}_{2L}^{(0)}\tilde{\varphi}\right]\tau_R^{(0)}-M\left[\bar{L}_{1L}^{(0)}L_{1R}^{(0)} +\bar{L}_{2L}^{(0)}L_{2R}^{(0)}\right]\nonumber\\
					   &-&\tilde{M}\left[\bar{L}_{3L}^{(0)}L_{3R}^{(0)}+\bar{Y}_{1L}^{(0)}Y_{1R}^{0}
%+\bar{N}_{2L}^{(0)}N_{2R}^{0}
\right]
-\tilde{y}\,\bar{l}_L^{(0)}\sigma^I\varphi L_{3R}^{(0)I} - \hat y \left[\bar{L}_{1L}^{(0)}\sigma^I \varphi-\bar{L}_{2L}^{(0)}\sigma^I \tilde{\varphi}\right]L_{3R}^{(0)I}\nonumber\\
					&-&\sqrt{2} \hat y 
%\left[\bar{L}_{1L}^{(0)}\tilde{\varphi} N_{2R}^{(0)}+
\bar{L}_{2L}^{(0)}\varphi Y_{1R}^{(0)}
%\right]
+\bar y^\ast \left[\bar{L}_{1R}^{(0)}\sigma^I \varphi-\bar{L}_{2R}^{(0)}\sigma^I \tilde{\varphi}\right]L_{3L}^{(0)I}
%\nonumber\\&+&
+\sqrt{2}\bar y^\ast 
%\left[\bar{L}_{1R}^{(0)}\varphi N_{2L}^{(0)}+
\bar{L}_{2R}^{(0)}\varphi Y_{1L}^{(0)}
%\right]-\sqrt 2 \tilde y\, \bar{l}_L^{(0)}\tilde\varphi N_{2R}^{(0)}-\sqrt 2\dot y \bar{L}_{1L}^{(0)}\tilde\varphi \nu_R^{(0)}
+\mathrm{h.c.}\,.
\end{eqnarray}

 After EWSB, we obtain the mass matrices 
\begin{eqnarray}
\label{eq:M101}
%	\mathcal{M}_N&=v &\begin{pmatrix} \frac{y_{\nu}}{\sqrt2} &0&\tilde{y}&-\tilde{y}\\ \dot{y} &\frac 1 v M_l& \hat y&- \hat y\\ 0 &-\bar{y}&\frac 1 v \tilde{M}&0\\0&\bar{y}&0&\frac 1 v \tilde{M}\end{pmatrix},\quad
\mathcal{M}_E=\frac{v}{\sqrt 2}\begin{pmatrix}
	y & 0 & 0 &-\tilde{y}\\ y^\prime & \frac{\sqrt2}{v}M & 0 &- \hat y\\ y^\prime & 0 & \frac{\sqrt 2}{v} M &- \hat y \\ 0&\bar{y}&\bar{y}&\frac{\sqrt 2}{v} \tilde{M}\end{pmatrix}, \qquad \mathcal{M}_Y= v \begin{pmatrix}\frac 1 v \tilde{M} & - \bar{y}&0\\ \hat y& \frac 1 v M&-\hat y\\ 0&\bar{y}&\frac 1 v \tilde{M} \end{pmatrix},
\end{eqnarray}
for the $Q=
%0,
-1,-2$ leptons, respectively.
Again, the natural size of the parameters appearing in (\ref{eq:M101}) can be motivated from the expected degree of compositeness of the contributing particles, determining the overlap and thus the mass-mixings or more general considerations. We will give the ranges of parameters that we employ in Section \ref{sec:expl}.

The rotations to the mass  basis will be in analogy to (\ref{eq:rot}), but now featuring larger matrices. 
We will resort to numerical methods for these diagonalizations in the following.
Note that, if we are only interested in sums of ratios of Higgs couplings over masses, we can arrive at simple analytical expressions, avoiding the diagonalization procedure, see Section~\ref{sec:Higgs}. 

The couplings of the fermions to the Higgs boson are now given by
\begin{equation}
\label{eq:gH10}
{\cal L}_h=\sum_{f=E,Y}\bar \Psi_L^{f10\,(0)} g_{h10}^{f(0)} \Psi_R^{f10\,(0)}\,  h+ {\rm h.c.} \,,
\end{equation}
where 
\begin{equation}
\begin{split}
%&\Psi^{N10\,(0)}\equiv (\nu^{(0)},N_1^{(0)},N_2^{(0)},N_3^{(0)})^T,\quad
\Psi^{E10\,(0)}\equiv (\tau^{(0)},E_1^{(0)},E_2^{(0)},E_3^{(0)})^T,\qquad
\Psi^{Y10\,(0)}\equiv (Y_1^{(0)},Y_2^{(0)},Y_3^{(0)})^T,
\end{split}
\end{equation}
 and
\begin{equation}
\label{eq:gh10}
g_{h10}^{f(0)}=\frac{\partial \mathcal{M}_f}{\partial v},
\end{equation}
with $f=E,Y$.
After rotating to the diagonal mass basis, the Higgs-coupling matrices become
\begin{equation}
\label{eq:ghmass}
g_{h10}^f= U_L^{f10\dagger} g_{h10}^{f(0)}  U_R^{f10}\,.
\end{equation}

\section{Higgs Production and Decay}
\label{sec:Higgs}

\subsection{General Structure}

\begin{figure}[!t]
	\begin{center}
	\mbox{\includegraphics[height=5.5cm]{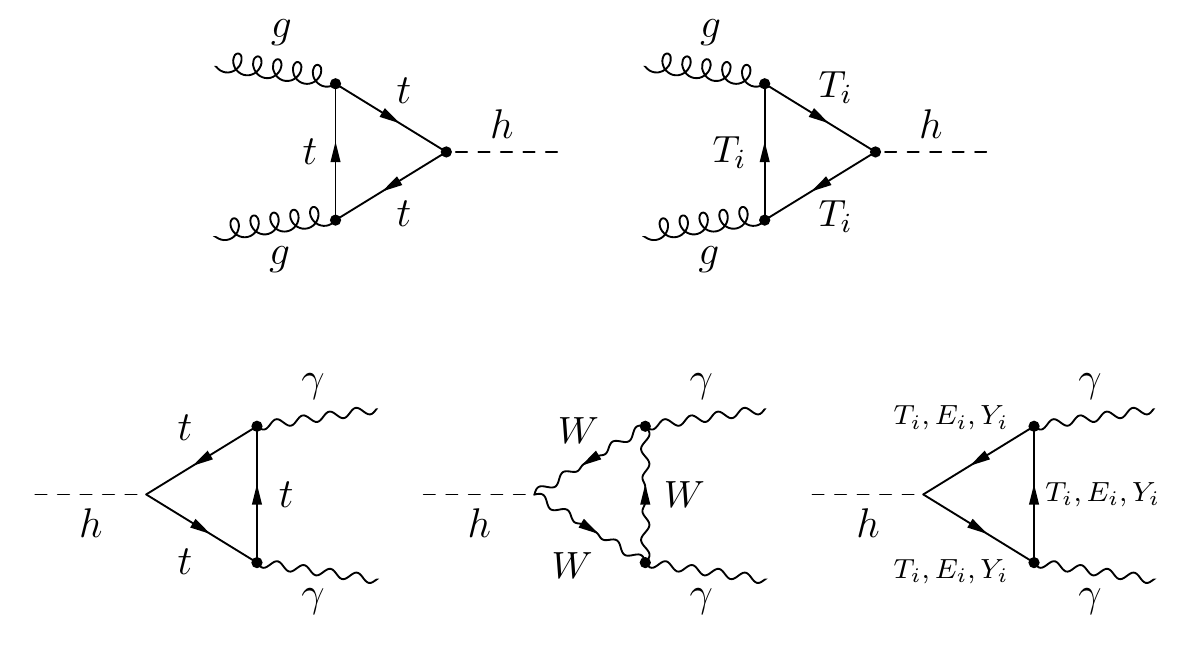}}
	\vspace{-1cm}
\parbox{15.5cm}{\caption{\label{fig:triangles}First row: Leading-order contribution to Higgs-boson production via gluon-gluon fusion and contribution from heavy quark resonances to the same process; Second row: Leading contributions to the Higgs decay into two photons, given by a top-quark loop and a $W^\pm$-boson loop, as well as contributions from heavy fermion resonances to the same process.}}
\end{center}
\vspace{0.3cm}
\end{figure}

The presence of the new resonances has significant implications on the production and decay of the Higgs boson, which will be worked out in this section. The most important production mechanism for the Higgs boson at hadron colliders is gluon-gluon fusion, which in the SM receives its main contribution from a top-quark triangle loop, with a large coupling to the Higgs, see the leftmost diagram in Figure \ref{fig:triangles}. In extensions of the SM this process can receive corrections from new colored particles that propagate in the loop (see the second diagram in the figure) as well as from modified couplings of the SM quarks in the loop to the Higgs boson. Both effects are present in the models we consider. We parametrize the corresponding deviations by a rescaling factor $\kappa_g^m$ as
\begin{equation}
\label{eq:gg}
\sigma(gg\to h)_{{\rm MCHM}_m}=|\kappa_g^m|^2\, \sigma(gg\to h)_{\rm SM}\,,
\end{equation}
whose explicit form will be given further below for $m=5,5$+$10$.
Subleading, but nevertheless important, channels for Higgs production at the LHC are vector-boson fusion (VBF) $qq^{(\prime)}\to qq^{(')} V^\ast V^\ast \to qq^{(\prime)}h$, with $V=W,Z$, associated vector-boson production $q\bar q^{(\prime)}\to V^\ast \to V h$,  and associated top-quark pair production $gg \to t \bar t^\ast t^\ast \bar t \to t \bar t h$, which all appear at the tree level, and the latter two will be abbreviated as $Vh$ and $tth$.\footnote{For the anatomy of these processes in the SM, see \cite{Djouadi:2005gi}.}
As the theories we consider only change the fermion sector of the SM, the tree-level couplings of the Higgs boson to weak gauge bosons remain standard-model like. The same is true for the couplings of the first two generations of fermions, which, due to their small masses, are assumed to have negligible mixings with the new resonances.\footnote{In extra dimensional extensions of the SM or composite Higgs models, with anarchic flavor structure, this assumption is motivated by the fact that the first two generations have negligible interactions with the KK excitations, or composite fermions.} Thus, to leading order,
\begin{eqnarray}
\label{eq:VB}
\sigma({\rm VBF})_{{\rm MCHM}_m}&=&\sigma({\rm VBF})_{\rm SM}\,,\\
\label{eq:VH}
\sigma(V h)_{{\rm MCHM}_m}&=&\sigma(V h)_{\rm SM}\,,
\end{eqnarray}
$m=5,5$+$10$. Since gauge invariance guarantees that the couplings of the fermions to the gluon and photon are unchanged,
the only correction to associated top-quark pair production in the models at hand arises through the deviation in the $h t\bar t$ vertex. The real part of such a coupling of SM-type fermions with the Higgs, $h \bar f f$, normalized to the SM, is given by
\begin{equation}
\label{eq:kappaf}
\kappa_f^5=v {\rm Re}\left[(g_{h5}^F)_{11}\right]/m_f\,,
\end{equation}
see (\ref{eq:gH5}), where $(f,F)=(t,T),(b,B),(\tau,E)$
and
\begin{equation}
\label{eq:kappaf10}
\kappa_f^{5+10} =  \begin{cases} v\, {\rm Re}\left[(g_{h10}^F)_{11}\right]/m_f\,, & (f,F)=(\tau,E)\\ 
	\kappa_f^5\,, & f=t,b \end{cases}\,,
\end{equation}
see (\ref{eq:gH10}).
We then get
\begin{equation}
\label{eq:tth}
\sigma(tth)_{{\rm MCHM}_m}=(\kappa_t^5)^2\, \sigma(tth)_{\rm SM}\,,
\end{equation}
$m=5,5$+$10$.

We now turn to the decays of the Higgs boson. The most important modes for a Higgs of $m_h \approx 125$\,GeV  are $h\to \gamma\gamma,WW,ZZ,bb,\tau\tau,gg$, where the last one is extremely difficult to measure. Moreover, the decays to two photons or gluons are loop processes, whereas the other decays happen at the tree level.
We parametrize deviations from the SM as
\begin{equation}
\label{eq:Gam}
\Gamma(h\to ff)_{{\rm MCHM}_m}=|\kappa_f^m|^2\, \Gamma(h\to ff)_{\rm SM}\,,
\end{equation}
$f=\gamma,W,Z,b,\tau,g$.
As discussed before, the decay to two vector bosons is unchanged in the models considered
\begin{equation}
\label{eq:WZ}
\kappa_W^m=\kappa_Z^m=1\,,
\end{equation}
$m=5,5$+$10$.
Beyond that, the rescaling factors for the (tree-level) decays into two fermions, entering (\ref{eq:Gam}),  have already been specified in (\ref{eq:kappaf}) and (\ref{eq:kappaf10}).

In the following, we will derive the explicit structure of the remaining rescaling factors, corresponding to loop precesses, which have not been detailed further yet, {\it i.e.}, $\kappa_g^m$ and $\kappa_\gamma^m$. Note that the first one enters in the same form in the gluon-gluon fusion process and in the decay of the Higgs to two gluons. Further below, we will relate the different rescalings to the parameters of the models under consideration.

For the effective coupling to gluons we arrive at
\begin{equation} \label{eq:kappag} 
  \kappa_g^5 = \kappa_g^{5+10} = \frac{{\displaystyle
      \sum}_{f = t, b} \, \kappa_f^5 \hspace{0.25mm} A_{q}^h
    (\tau_f)\, + \nu_T^5 }{ {\displaystyle \sum}_{f = t,
      b} \; A_{q}^h (\tau_f)} \,,
\end{equation}
where $\tau_f \equiv 4 \hspace{0,25mm} m_f^2/m_h^2$. This expression is valid for both models, since we did not modify the quark sector in the MCHM$_{5+10}$ with respect to the MCHM$_5$.
The first term in the numerator takes into account the change in the $h \bar ff$ vertices appearing in the triangle loop, where we kept the contributions from the top and the bottom quark (see upper-left diagram of Figure \ref{fig:triangles} for $f=t$). The corresponding loop function $A_{q}^h (\tau_f)$ approaches $1$ for $\tau_f \to \infty$ (which is already a good approximation for $\tau_t\approx 7.5$, leading to  $A_{q}^h (\tau_t)\approx 1.03$) and vanishes proportional to $\tau_f$ for $\tau_f \to 0$. Its analytic form is given in
Appendix~\ref{app:formfactors}. We again consider only third generation fermions as the couplings of the others to the Higgs boson are strongly suppressed. The second term in the numerator of (\ref{eq:kappag}) represents the contribution arising from the virtual exchange of the heavy vector quarks (the top custodians), contained in $\Psi^{T5}$, which have significant couplings to the Higgs, see the second diagram in Figure \ref{fig:triangles}. Remember that these resonances all couple diagonal to the gluons. Introducing already the corresponding lepton quantities which do not enter (\ref{eq:kappag}), but will be needed later, one obtains
\begin{equation}\label{eq:nu5}
    \nu_F^5 =  \begin{cases} \displaystyle \,v \;\sum_{n = 2}^3 \; \frac{{\rm Re}\left[ (g_{h5}^F)_{nn}\right]}{m_{F_{n-1}}} \,, & F=T,E\\ 
	\hspace{1.8cm} 0 \hspace{1.3cm} \,, & F=Y \qquad , \end{cases}
\end{equation}
\begin{equation}\label{eq:nu10}
    \nu_F^{5+10} =  \begin{cases} \displaystyle \,v \;\sum_{n = 2}^4 \; \frac{{\rm Re}\left[ (g_{h10}^F)_{nn}\right]}{m_{F_{n-1}}} \,, & F=E\\ 
	\displaystyle \,v \;\sum_{n = 1}^3 \;  \frac{{\rm Re}\left[ (g_{h10}^F)_{nn}\right]}{m_{F_n}} \,, & F=Y \qquad. \end{cases}
\end{equation}
Note that since all the new resonances are much heavier than the Higgs boson, the loop functions that would multiply the above quantities are equal to 1 to excellent approximation and thus could be omitted.

For the effective coupling to photons we obtain
\begin{equation} \label{eq:kappagamma} 
\kappa_\gamma^m =
  \frac{{\displaystyle \sum}_{f = t, b} \; N_c \hspace{0.5mm} Q_f^2
    \hspace{0.5mm} \kappa_f^m \hspace{0.25mm} A_{q}^h (\tau_f) + Q_\tau^2 
    \kappa_\tau^m \hspace{0.25mm} A_{q}^h (\tau_\tau) +
    A_W^h (\tau_W) +  N_c \hspace{0.5mm} Q_t^2
    \hspace{0.5mm} \nu_T^5+ {\displaystyle \sum}_{F = E, Y} Q_F^2
    \hspace{0.5mm} \nu_F^m }{{\displaystyle \sum}_{i = t,
      b} \; N_c \hspace{0.5mm}  Q_i^2 \hspace{0.5mm} A_{q}^h (\tau_i) 
    +Q_\tau^2 \hspace{0.5mm} A_{q}^h (\tau_\tau) +
    A_W^h (\tau_W)}\,,
\end{equation}
where $N_c =3$, $Q_t = 2/3$, $Q_b =
-1/3$, $Q_{\tau,E} = -1$, $Q_Y =
-2$, and $\tau_W \equiv 4 \hspace{0.25mm} m_W^2/
m_h^2$. Here, we have already employed that $\kappa_W^{5,\,5+10}=1$.
The explicit expression for the form factor $A_{W}^h
(\tau_W)$, encoding the $W^\pm$-boson contribution, can be found in
Appendix~\ref{app:formfactors}. The other quantities entering (\ref{eq:kappagamma}) have already been given before, see (\ref{eq:kappaf}), (\ref{eq:kappaf10}), (\ref{eq:nu5}) and (\ref{eq:nu10}). The first, second, and third terms in
the numerator above describe the effects of virtual SM-type quark, lepton, and $W^\pm$-boson
exchange, respectively. The fourth and fifth term, on the other hand, correspond to the contributions of the custodians. Examples for corresponding one-loop graphs are shown in the second row of Figure \ref{fig:triangles}. Note that the amplitude proportional to $A_{W}^h (\tau_W)\approx -6.25$ dominates in the SM and interferes destructively with the fermion contribution $A_{q}^h (\tau_f)$. Thus, adding just SM-like fermions, like a chiral $t^\prime$, will reduce the effective coupling to photons. However, if the new fermions get part of their masses from another mechanism than the Higgs, like vector-like quarks or leptons, it is in principle possible to enhance $\kappa_\gamma^m$. We will see, that this is indeed the case for the MCHM$_{5+10}$.

\subsection{Explicit Results in the Models at Hand}
\label{sec:expl}

We now give the explicit predictions for the various quantities defined in the previous section for both the MCHM$_5$ and the MCHM$_{5+10}$. To that extend, we should also specify the values we use for the free parameters of the models, which we will do further below. 

\subsubsection{MCHM$_5$}

\label{sec:MCHM5Higgs}
Let us start with analyzing the MCHM$_{5}$, where we obtained easy analytic formulas for the masses and Higgs couplings in Section \ref{sec:MCHM5}. Employing (\ref{eq:gh5}) we directly arrive at
\begin{eqnarray}
\label{eq:kappatau5}
\kappa_\tau^5=(c_R^\tau)^2\,,\\
\label{eq:kappt5}
\kappa_t^5=(c_R^t)^2\,,
\end{eqnarray}
while
\begin{equation}
\kappa_b^5=1\,.
\end{equation}
The Higgs couplings to two $\tau$ leptons and two top quarks are thus predicted to be reduced in the MCHM$_5$. Note that at this point $c_R^{\tau,t}$ are free parameters of the model and thus allowed to take any value in their range of definition $0\leq c_R^{\tau,t} \leq 1$. Physically, these parameters describe the mixings of the $t_R$ and the $\tau_R$ to the new physics, which can reach from the decoupling limit $ c_R^{\tau,t}\to 1$ up to a {\cal O}(1) mixing, which would start at $ c_R^{\tau,t} \sim 1/\sqrt 2$. For smaller values of $ c_R^{\tau,t}$, the $\tau$ and the top quark will have typically stronger couplings with the new physics than the generic couplings (vector like masses) within the new physics sector.
From the perspective of a composite Higgs model, $ c_R^{\tau,t} \leq 1/\sqrt 2$ would correspond to a full compositeness, which should be kept in mind when studying the impact on Higgs decays. In that context, note that for $M\gg m_h$, which we will assume in the following, the predictions in Higgs physics are in principle independent of the vector-like mass $M$ itself. However, if direct searches push $M$ beyond the TeV range (see (\ref{eq:ME})), the scale of the absolute mass-mixing between elementaries and composites that is needed for $c_R^{\tau,t}\ll 1$ could become problematic.\footnote{Note that in the full 5D/composite Higgs models there is a correlation between the elementary-composite mixing and the mass of the 
light custodians. However, in the lepton sector the direct bounds for 
current luminosities are weak, see e.g. \cite{delAguila:2010es}, and do 
not affect significantly the parameter space. Regarding the quark 
sector, the latest and most stringent direct production bounds on the 
masses of vector-like quarks not coupling to the light generations are 
about $\sim 700~\GeV$ \cite{ATLASvlq}, however assuming a Higgs with 
SM couplings to fermions and gauge bosons. Anyway, all observables 
studied in the following are to excellent approximation independent of 
this parameter, due to a cancellation of the contributions of the top 
resonances and the SM-like top quark, see below. The only exception is 
the tree-level $ht\bar{t}$ coupling, which could be more constrained, due to the direct bound, 
leading to a reduced effect in $h \to b\bar{b}$ via $tth$ production, 
see Figure~\ref{fig:gamtaub1}.}

For the quantities related to the couplings of the heavy resonances, see (\ref{eq:nu5}), we obtain again from (\ref{eq:gh5})
\begin{eqnarray}
\label{eq:nu5s}
\nu_E^5=(s_R^\tau)^2\,,\\
\nu_T^5=(s_R^t)^2\,.
\end{eqnarray}

Combining these results, we arrive at
\begin{equation}  
\label{eq:kappag5}
  \kappa_g^5 \approx \frac{(c_R^t)^2 \hspace{0.25mm}+ A_{q}^h(\tau_b)\, 
     + (s_R^t)^2}{ 1+A_{q}^h (\tau_b)}  = 1\,,
\end{equation}
where we have used $A_{q}^h(\tau_t)\approx 1$. Thus, neglecting small deviations from this approximation,
the production cross section for the Higgs boson in gluon-gluon fusion is unchanged in the MCHM$_{5}$, if one considers the low energy model of this work. There is a cancellation between corrections to the top Yukawa coupling and the contributions of the new top resonances, leading to a total contribution (normalized to the SM) of $(c_R^t)^2 + (s_R^t)^2 =1$, which is independent of the parameters of the fermion sector. This result agrees with the findings of \cite{Falkowski:2007hz} (see also \cite{Azatov:2011qy,Furlan:2011uq}), which considers a complete composite Higgs model and thus additionally takes into account effects of the non-linearity of the Higgs sector, which are suppressed by $v^2/f^2$.
For the effective coupling of the Higgs to two photons we obtain in the same way
\begin{eqnarray}
\label{eq:kappagam5}
\kappa_\gamma^5 &=&
  \frac{N_c (Q_t^2 (c_R^t)^2+Q_b^2
    \hspace{0.5mm} A_{q}^h (\tau_b) )+ Q_\tau^2 
    (c_R^\tau)^2 \hspace{0.25mm} A_{q}^h (\tau_\tau) +
    A_W^h (\tau_W) +  N_c \hspace{0.5mm} Q_t^2
    \hspace{0.5mm} (s_R^t)^2+  Q_\tau^2
    \hspace{0.5mm} (s_R^\tau)^2}{N_c (Q_t^2+Q_b^2 \hspace{0.5mm} A_{q}^h (\tau_b) )
   +  Q_\tau^2 \hspace{0.5mm} A_{q}^h (\tau_\tau)  +  A_W^h (\tau_W)}\nonumber\\
  &=&
  \frac{N_c (Q_t^2 +Q_b^2
    \hspace{0.5mm} A_{q}^h (\tau_b) )+ Q_\tau^2 (
    (c_R^\tau)^2 \hspace{0.25mm} A_{q}^h (\tau_\tau) +  
     (s_R^\tau)^2)  +
    A_W^h (\tau_W) }{N_c (Q_t^2+Q_b^2 \hspace{0.5mm} A_{q}^h (\tau_b) )
   +  Q_\tau^2 \hspace{0.5mm} A_{q}^h (\tau_\tau)  +  A_W^h (\tau_W)}\,.
\end{eqnarray}
From the second line above, one can clearly see that due to the contribution of the new leptons, the structure that lead to $\kappa_g^5\approx 1$ (or $\kappa_g^5\approx (1-2 v^2/f^2)/\sqrt{1-v^2/f^2}$  in full composite models, see Section \ref{sec:vf}) for the Higgs coupling to two gluons is broken in $\kappa_\gamma^5$. Taking into account the lepton sector introduces a pattern which has not been considered in \cite{Falkowski:2007hz}, namely a light particle ($m_\tau\ll m_h$) with a significant composite component (due to the mechanism which generates the light custodians). Because of $|A_{q}^h (\tau_\tau)|\approx0.02\ll1$, the contributions of the SM-type lepton and the corresponding heavy resonances do not add up to a result which is independent from the model parameters. Their effect does {\it not} cancel, which is a very interesting and distinct feature of the lepton sector in these composite Higgs models.

Moreover, neglecting the tiny contribution proportional to $A_{q}^h (\tau_\tau) $, we can easily see that
\begin{equation}
\label{eq:kappagam5ap}
\kappa_\gamma^5 \approx
  \frac{-5 + 
     (s_R^\tau)^2}{-5}\,.
\end{equation}
Thus, using the fact that $0<(s_R^\tau)^2<1$ in the MCHM$_5$, we arrive at the clear prediction $\kappa_\gamma^5<1$.
Physically, this is due to the fact that the new vector-like lepton adds a positive contribution to the numerator of (\ref{eq:kappagam5}), interfering destructively with the leading term proportional to $A_W^h (\tau_W)$.
Note that, if we want to think of a complete GHU model producing our setup, we are missing 4 (6) heavier vector-like resonances of the first KK level in the T,E (B) sector which couple to the Higgs boson and could potentially enter the low energy predictions. However, we have seen that the light top-like modes present in our setup already contribute a structure $ \sim (c_R^t)^2 + (s_R^t)^2 =1$ to $\kappa_{g,\gamma}^5$, independent of the fermion parameters, and similar, but featuring different loop functions, for the $\tau$. This agrees with the result of the corresponding full model, see \cite{Falkowski:2007hz,Azatov:2011qy} (neglecting $v/f$ corrections due to the modification of the Higgs sector for the moment). Thus the full fermion structure of the 5D model is present in our simplified low energy setup of light custodians, parametrized by $c_R$ and $s_R$ \cite{next} (if one wants to consider the setup as an effective theory of a composite model). For the B sector, an analogous discussion holds, since neglecting $v/f$ corrections, the predictions are unchanged with respect to the SM in our setup.
Concerning potential contributions of resonances belonging to the first two generations, keep in mind that in GHU the total changes in Higgs physics due to light generations is negligible \cite{Falkowski:2007hz}.
We will comment on further corrections to the above picture in full GHU models later on.

To summarize the findings of this section, the Higgs-production cross section is unchanged in our MCHM$_5$ setup in the gluon-gluon fusion, VBF and associated $W^\pm$-production channels to very good approximation, see (\ref{eq:gg})-(\ref{eq:VH}) and (\ref{eq:kappag5}), whereas it is reduced in associated top-quark pair production, see (\ref{eq:tth}) and (\ref{eq:kappt5}). As discussed above, the decay cross sections into
photons, $\tau$ leptons and top quarks are reduced in the model. The explicit results for the production cross section times branching fraction in the different channels will be discussed in detail in Section \ref{sec:pheno}.

\subsubsection{MCHM$_{5+10}$}

\label{sec:Higgs10}
We now move over to the case of the MCHM$_{5+10}$.
Here, naively it seems that one would have to resort completely to numerical methods to diagonalize the more complicated mass matrices  (\ref{eq:M101}) to finally obtain the Higgs couplings in the mass basis (\ref{eq:ghmass}) that enter the various effective couplings. 
However, we can use a trick to avoid this procedure. First, note that, neglecting the loop functions,
(\ref{eq:kappagamma}) contains the structure
\begin{equation}
\label{eq:lam}
\lambda_F\equiv 
 \begin{cases}\kappa_{\tau}^{5+10}+\nu_F^{5+10}= v \displaystyle \sum_{n = 1}^4 \; \frac{{\rm Re}\left[ (g_{h10}^F)_{nn}\right]}{m_{F_{n-1}}}\,, & F=E\\ 
	\nu_F^{5+10}= v \displaystyle \sum_{n = 1}^3 \; \frac{{\rm Re}\left[ (g_{h10}^F)_{nn}\right]}{m_{F_{n}}} \,, & F=Y \quad, \end{cases}
\end{equation}
where, $m_{E_0}=m_\tau$.
It turns out that these expressions can be summed analytically in closed form by using the relation (see e.g. \cite{Falkowski:2007hz})
\begin{equation}
\label{eq:magic}
\sum_n \; \frac{(g_{h10}^F)_{nn}}{\bar m_{F_{n}}}= \frac{\partial\log(\det {\cal M}_F)}{\partial v}\,,
\end{equation}
where $\bar m_{F_{n}}$ denote the mass eigenvalues belonging to the mass matrix ${\cal M}_F$, {\it i.e.}
$\bar m_{F_{n}}=m_{F_{n-1}}$ for $F=E$ and $\bar m_{F_{n}}=m_{F_n}$ for $F=Y$.
Note that in order to evaluate the right hand side of (\ref{eq:magic}) one does not have to go to the mass basis. This allows us directly to also arrive at analytical expressions for the MCHM$_{5+10}$. Applying (\ref{eq:magic}) to (\ref{eq:M101}) we obtain\footnote{In the same way we could have derived the corresponding expressions for the MCHM$_5$, getting the same results as the ones obtained employing the mass basis.}

\begin{eqnarray}
        \frac{\partial \log (\mathrm{det}\mathcal{M}_{E})}{\partial v}&=&\frac{1}{v}\frac{3 v^2\, y\bar y \hat y+y M\tilde{M}-3v^2\, \bar{y} y^\prime \tilde{y}}{v^2\, y\bar y \hat y+y M\tilde{M}-v^2\, \bar{y} y^\prime \tilde{y}}\nonumber\\
                                                                                                                                                                                                                                                                                                                                                                                                                                                                                                                   &=&\frac{1}{v}\left(1+2 v^2 \frac{\bar{y} \hat y}{M \tilde M}-2 v^2 \frac{\tilde y \bar{y} y^\prime}{y M \tilde{M}}+\mathcal{O}(\epsilon^3)\right),
        \label{det:MET}\\
        \frac{\partial \log (\mathrm{det}\mathcal{M}_{Y})}{\partial v}&=&\frac{1}{v}\frac{4v^2 \bar y \hat y}{M \tilde{M} + 2 v^2 \bar{y} \hat y}=\frac{1}{v}\left(4v^2\, \frac{\bar{y} \hat y}{M \tilde{M}}+\mathcal{O}(\epsilon^3)\right),\label{det:MYA}
\end{eqnarray}
where we have denoted $\epsilon \sim v/M,v/\tilde{M}$. 
Note that, for a sector that involves only heavy particles, {\it i.e.}, $F=Y$ (where $m_{Y_i} \gg m_h$ leads to $A_{q}^h (\tau)\approx 1$) the structure $\lambda_Y$ (\ref{eq:lam}) can directly be found in (\ref{eq:kappagamma}).

The structure leading to the expression (\ref{det:MET}) is however broken since the corresponding sector involves light SM-fermions, with a different loop function. It is nevertheless possible to extract the light-mode contribution from the corresponding equations to order $\epsilon^2$ through the dimension six effective Lagrangian obtained from the integration of the corresponding vector-like fermions. Following \cite{delAguila:2000rc} we obtain
\begin{eqnarray}
        v\frac{(g_{h10}^{E})_{11}}{m_{E_0}}&=&1-v^2\left(\frac{|\tilde{y}|^2}{2 \tilde{M}^2}+\frac{|y^\prime|^2}{M^2}+2\frac{\tilde{y}\bar{y} y^\prime}{y M \tilde{M}}\right) +\mathcal{O}(\epsilon^3),
\label{eq:0mode}\\
       v\sum_{n=2}^4     \frac{(g_{h10}^{E})_{nn}}{m_{E_{n-1}}}&=& v^2 \left( \frac{|y|^2}{2\tilde{M}^2}+\frac{|y^\prime|^2}{M^2}+2\frac{\bar{y} \hat y}{M \tilde{M}}\right) +\mathcal{O}(\epsilon^3)\, .
\label{eq:KKsum}
\end{eqnarray}
In this approximation it is thus possible to use (\ref{eq:magic})-(\ref{eq:KKsum}) to sum up the different contributions to  $\kappa_\gamma^{5+10}$ in closed form to analytical results, see (\ref{eq:kappaf10}), (\ref{eq:nu10}), and (\ref{eq:kappagamma}). At this point some comments are in order.

\begin{table}
\begin{center}
\begin{tabular}{|c|c|}
\hline
Parameter & central value [GeV]\\
\hline \hline
$ |m \approx - \tilde m|$ & 1  \\
   \hline
$ |m^\prime  \approx -  \hat m| $  & 100 \\
   \hline
$ |\bar m| $  & 100 \\
   \hline
$ M $  & 400  \\
   \hline
$ \tilde M $  & 450\\
   \hline
\end{tabular} \quad 
\end{center}
\vspace{-3.5mm}
\parbox{15.5cm}{\caption{\label{tab:paras}Assumptions for the free parameters of the MCHM$_{5+10}$, defining $m = v/\sqrt 2 \, y$  for the various Yukawa couplings. All values are varied around the central value $m_{\rm cent}$ in the range $[0.4,2.5] m_{\rm cent}$. Moreover, besides the vector-like mass terms $M$ and $\tilde M$, all parameters are allowed to have arbitrary phases.}}
\end{table}

The expressions above contain quite a number of parameters and so will  $\kappa_i^{5+10}$, $i=\tau,\gamma$. In order to make transparent the predictions of the MCHM$_{5+10}$, we should specify the assumptions we make on the parameters and visualize the predictions for the different quantities entering the effective Higgs couplings. To that extend we scan over the parameter-space of the model, varying the mass parameters in the range $[0.4,2.5] m_{\rm cent}$ around their central values $m_{\rm cent}$, which are given in Table \ref{tab:paras}, with a flat distribution (for the various Yukawa couplings we define $m = v/\sqrt 2 \,y$).
 Note that all parameters, besides the vector-like mass terms $M$ and $\tilde M$ are allowed to have arbitrary phases. The magnitude of the corresponding parameters is motivated by the assumption that only the $\tau_R$, responsible for the relatively large mass of the $\tau$, couples significantly to the new physics. As discussed, this corresponds for example to the low energy tail of composite models/GHU models featuring an $A_4$ symmetry, and matches well with phenomenological constraints. The chosen range for the (Kaluza-Klein)  masses of the light resonances $160\,{\rm GeV}< M,\tilde M <1125$\,GeV corresponds to the natural range of models addressing the gauge hierarchy problem.
The parameters will be constrained to result in a mass for the $\tau$ lepton that is in agreement with the experimental value, evaluated at the scale of the new resonances.
%Studying the changes to the CKM matrix, which are suppressed by $v/m_{\rm cust}$, requires a specification of the details concerning the first two families, which is beyond the scope of this work. 
As it turns out that the $\tau_R$ has a similar degree of compositeness as its light custodian partners (with opposite ``sign''), we assume that $\tilde m=-m(1\pm 10 \%)$ and $\hat m=- m^\prime(1\pm 10 \%)$. This approximate equality has important implications on the structure of (\ref{det:MET}), which then becomes $\partial \log (\mathrm{det}\mathcal{M}_E)/\partial v\approx 1/v$. Thus, as in the case of the MCHM$_5$, there is a cancellation between the correction to the SM $\tau$ Yukawa and the heavy resonances entering $\kappa_\gamma^m$, that would lead to the same contribution of the complete $\tau$ sector as the $\tau$ contribution in the SM, {\it i.e.} a cancellation of the new physics effects {\it if} the $\tau$ was heavier than the Higgs boson. However, as for the MCHM$_5$, this is broken completely by the different loop functions for the $\tau$ and its custodian partners. Nevertheless, this structure assures that, like in the MCHM$_5$, we also capture the whole physics of the complete KK tower, if we want to consider our setup as a low energy tail of a GHU model.\footnote{Note that the structure leading to $\partial 
\log(\mathrm{det}\mathcal{M}_E)/\partial v\approx1/v$ grasps the physics of the 
whole KK tower even in the case of the MCHM$_{5+10}$, where due to the various 
fermion representations present in the model two different trigonometric 
functions can arise \cite{Azatov:2011qy}. This is due to the fact that the more 
composite $\tau_{R}$ just can mix with the \textbf{10}, being only the 
almost elementary SM doublet $l_L$ who connects both sectors. Therefore, 
similarly to what happens in the model considered in Appendix B of 
\cite{Azatov:2011qy}, the breaking of the above pattern in the complete 
composite model will be governed by the magnitude of $\lambda_l^{(10)}$, a 
small parameter controlled by the size of the tau mass over the $\tau_R$ 
linear coupling.} In contrast to the case of the MCHM$_5$, there are now additional contributions from exotic $Q=-2$ fermions $Y$. In this case, as mentioned before, we do consider a complete KK level, as the missing heavy $Y$-fields belong to fundamental representations of $SO(5)$ and thus have negligible Higgs-couplings. 

We thus expect in principle quite different signatures in Higgs Physics in the MCHM$_{5+10}$ with respect to the MCHM$_5$.
Putting more SM-leptons or the quarks into a \textbf{10} of $SO(5)$ would spoil the above considerations and thus the model we consider is a conservative choice, if one wants to capture the full structure of a composite model via the light-particle spectrum. While we expect the numerical results to change once we put the whole third generation into a \textbf{10}, the {\it qualitative} behavior is expected nevertheless to be similar to our setup \cite{next} and thus the model can be seen as a simple setup that allows to understand the behavior of the lepton sector of the full MCHM$_{10}$.

\begin{figure}[!t]
\begin{center}
 \includegraphics[height=2.7in]{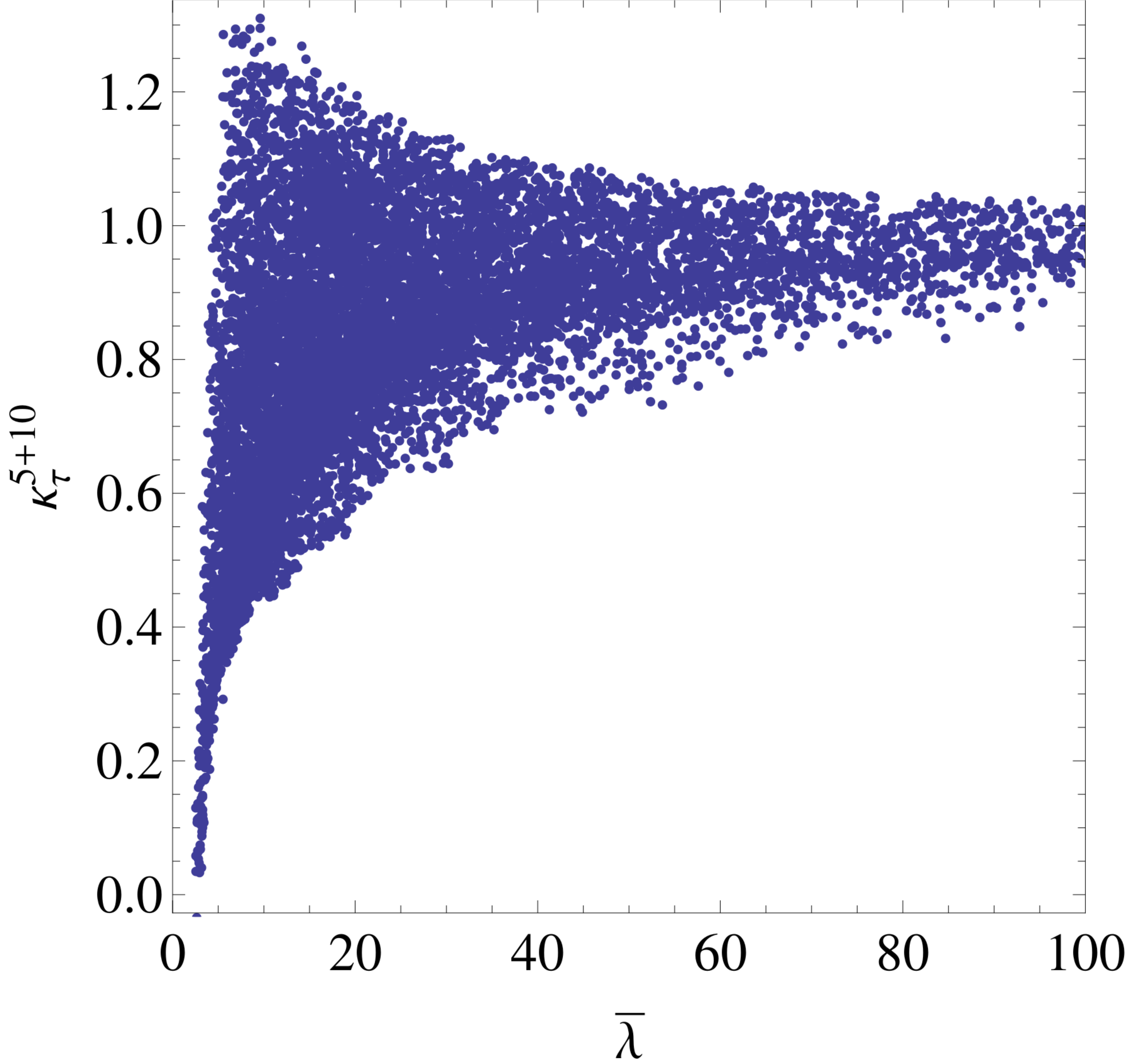}\quad
\includegraphics[height=2.7in]{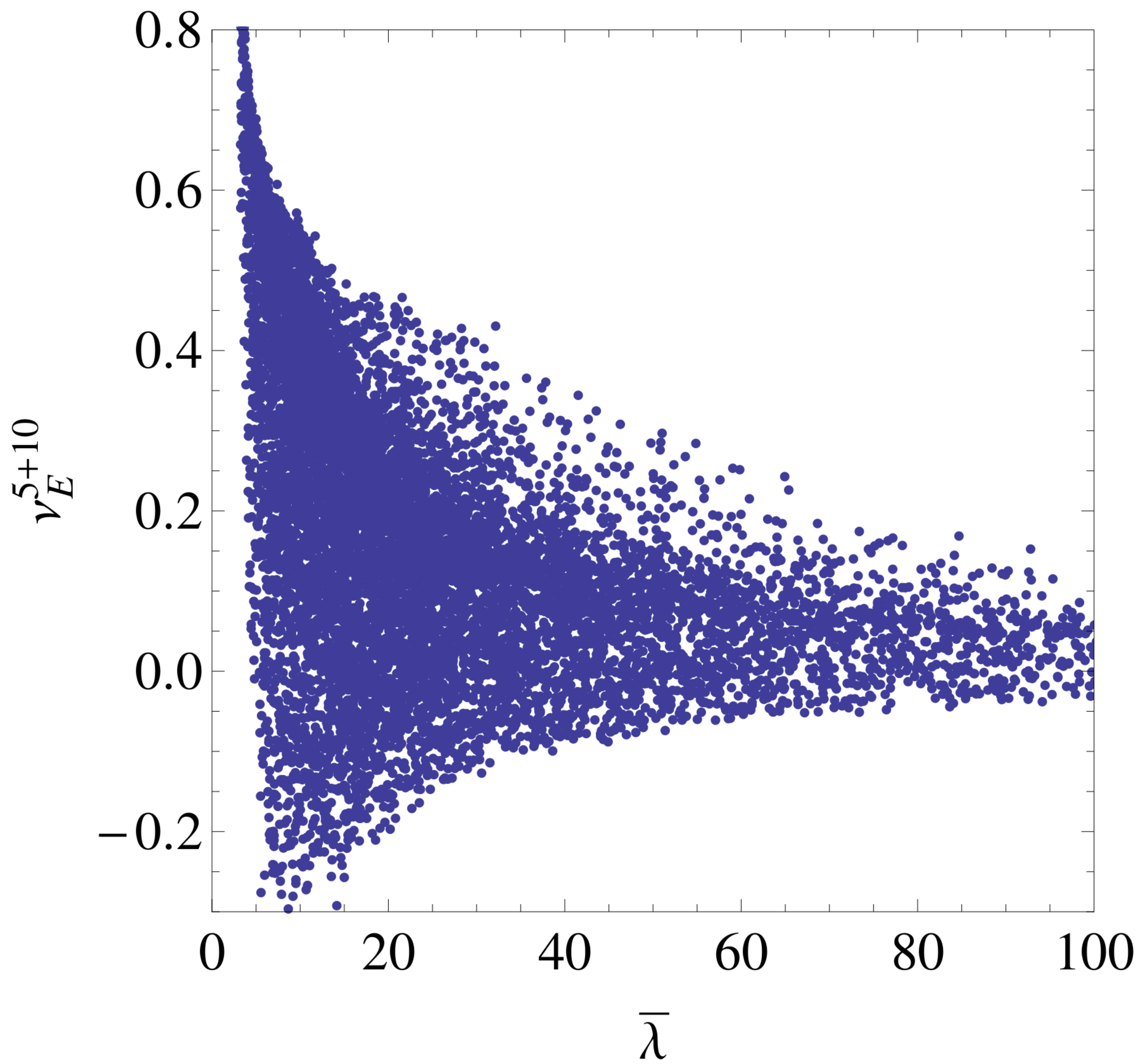}
\parbox{15.5cm}{\caption{\label{fig:kapnu} Predictions for the quantities $\kappa_\tau^{5+10}$ and $\nu_E^{5+10}$ in the MCHM$_{5+10}$ plotted versus the parameter $\bar \lambda=2 M \tilde M/(v^2 |y^\prime \bar y|)$, measuring the vector-like masses over the product of the compositeness of the right handed $\tau$ with the Yukawa couplings of the new resonances. The points correspond to a scan over the parameter-space of the model. See text for details.}}
\end{center}
\end{figure}

In order to explore the new features of the MCHM$_{5+10}$ we now show scatter plots representing the predictions for $\kappa_\tau^{5+10}$, $\nu_{E,Y}^{5+10}$, and $\lambda_E$ for 10000 parameter points. They have all been obtained by an exact numerical diagonalization of the mass matrices (\ref{eq:M101}) and a subsequent numerical evaluation of the Higgs couplings in the mass basis (\ref{eq:ghmass}), employing the parameters given in Table  \ref{tab:paras}, which are varied as described before.
We plot the results with respect to the parameter
\begin{equation}
\label{eq:lamb}
\bar \lambda=\frac{2 M \tilde M}{ v^2  |y^\prime \bar y|}\,, 
\end{equation}
which measures the vector-like masses over the compositeness of the right handed $\tau$ times the scale of the Yukawa couplings of the new resonances. Note that, due to $y^\prime \approx - \hat y$, at the same time it also is a measure for the size of the vector-like masses over the scale of the Yukawa couplings of the new resonances alone. It is thus expected that all the physical predictions of the model should scale with this parameter.

\begin{figure}[!t]
\begin{center}
\includegraphics[height=2.7in]{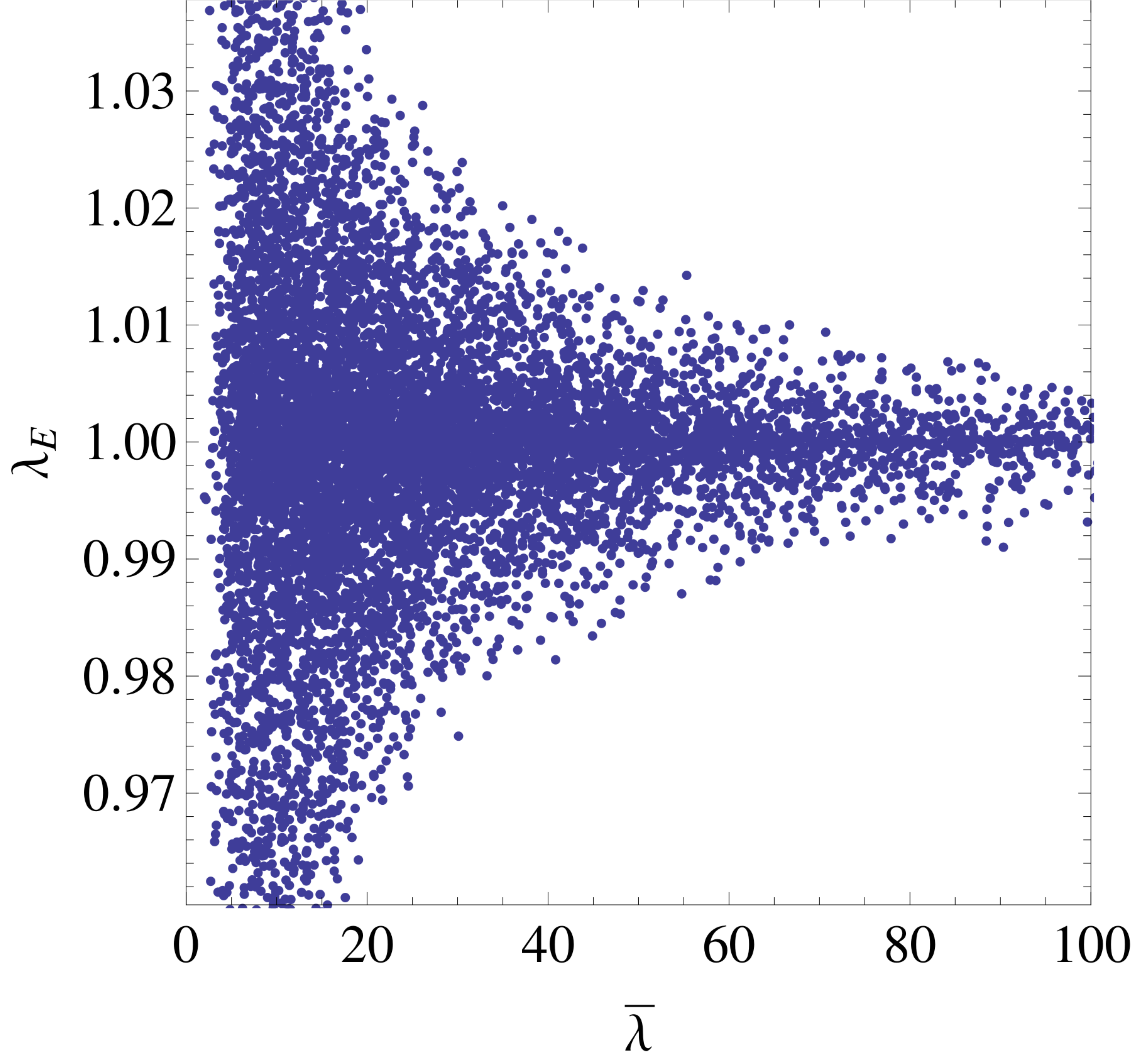}\quad
\includegraphics[height=2.7in]{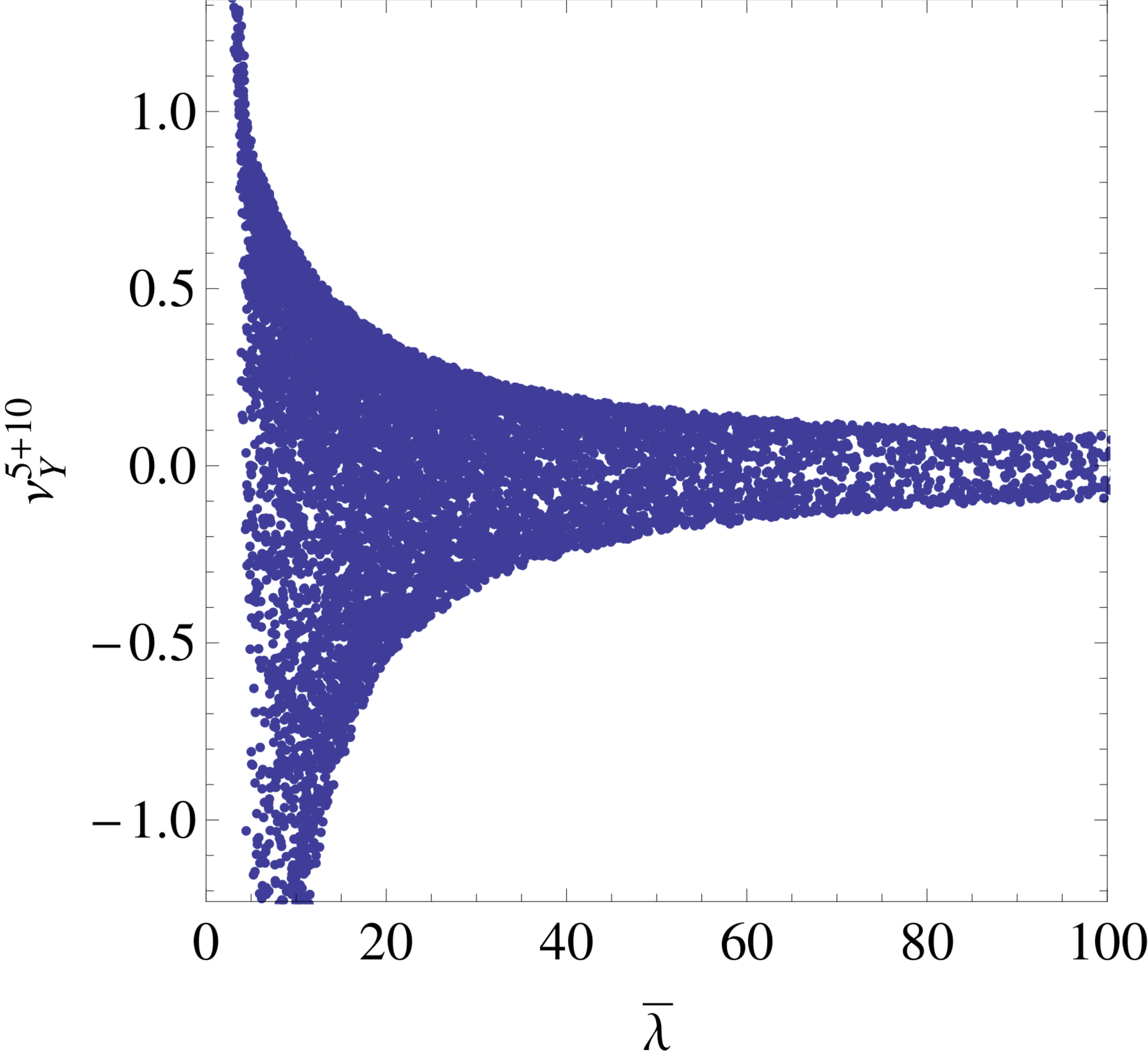}
\parbox{15.5cm}{\caption{\label{fig:nulam} Predictions for $\lambda_E$ and $\nu_Y^{5+10}$ in the MCHM$_{5+10}$ plotted versus $\bar \lambda$, measuring the scale of the vector-like masses of the new resonances over their Yukawa couplings.
The points correspond to a scan over the parameter-space of the model. See text for details.}}
\end{center}
\end{figure}

In Figure \ref{fig:kapnu} we show the effective coupling $\kappa_\tau^{5+10}$ as well as $\nu_E^{5+10}$ in the  MCHM$_{5+10}$ with respect to $\bar \lambda$. One can clearly see that also in the MCHM$_{5+10}$ the couplings of the $\tau$ lepton to the Higgs boson are generically reduced. They are mostly in the range $0<\kappa_{\tau}^{5+10}<1$, as for the MCHM$_5$ (see (\ref{eq:kappatau5})), however a small enhancement seems also possible. For a squared compositeness scale of roughly one order of magnitude below the squared scale of the light resonances, $\bar \lambda \sim 10$, one can have a depletion of up to $\kappa_\tau^{5+10} \sim 0.5$. In the decoupling regime of large $\bar \lambda$, one approaches the SM value of  $\kappa_\tau = 1$, as expected. The contributions of the resonances of the $\tau$ sector also become important for low $\bar \lambda$, {\it i.e.} of the order $\nu_E^{5+10}\sim 0.5$ for $\bar \lambda \sim 10$ (for the region with the largest density of scatter points), as can be read off from the right plot in the figure. From the plots one can already suspect the numerical confirmation of the discussion below  (\ref{det:MYA}), {\it i.e.}, that in the MCHM$_{5+10}$, in analogy to the MCHM$_5$, the relation $\lambda_E=\kappa_\tau^{5+10} + \nu_E^{5+10}=1$ holds to good approximation. This can be seen more clearly in the left panel of Figure~\ref{fig:nulam}, where we plot $\lambda_E$ versus $\bar \lambda$ and find that indeed $\lambda_E\approx1$. The breaking of the exact relation is due to the fact that we did allow for small variations $\tilde m=-m(1\pm 10 \%)$ and $\hat m=- m^\prime(1\pm 10 \%)$.
 Remember however, that both contributions to $\lambda_E$ enter $\kappa_\gamma^{5+10}$ with a different loop function.
In the right panel of Figure~\ref{fig:nulam}  we show our  predictions for $\nu_Y^{5+10}$, plotted again versus $\bar \lambda$. We can see that this contribution can become negative which allows for a constructive interference with the $W^\pm$-loop in $h\to \gamma\gamma$ and finally leads to a possible enhancement in the Higgs decay into two photons, which is not possible to get in the MCHM$_5$.

\begin{figure}[!t]
\begin{center}
\includegraphics[height=2.8in]{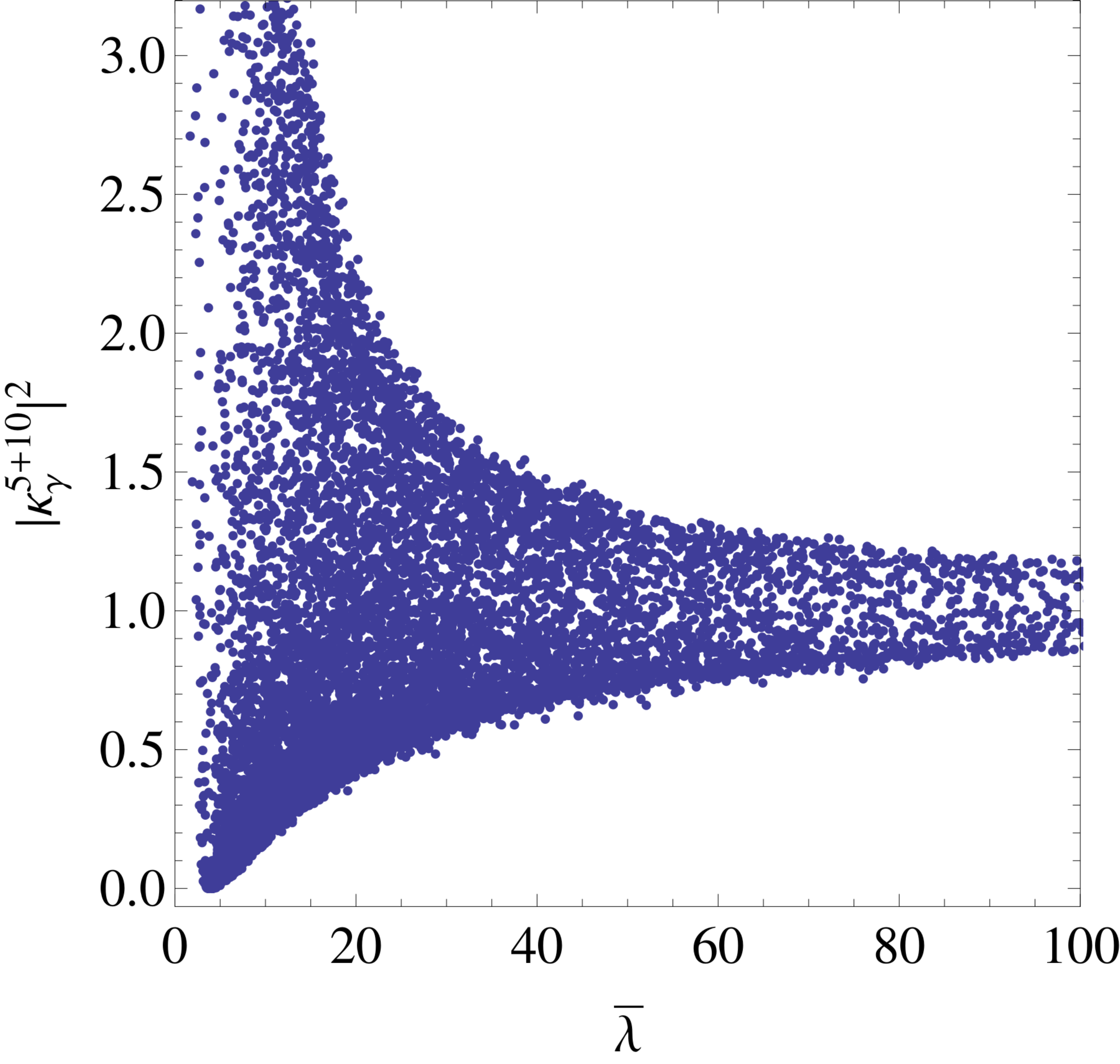}
\parbox{15.5cm}{\caption{\label{fig:kgam} Predictions for $|\kappa_{\gamma}^{5+10}|^2$  plotted versus $\bar \lambda$. Note that the dependence on the parameters of the quark sector drops out to good approximation. The points correspond to a scan over the parameter-space of the model. See text for details.}}
\end{center}
\end{figure}

Finally, we come to the results for $\kappa_{\gamma}^{5+10}$, which due to the fact that the quark sector is unchanged with respect to the MCHM$_5$ takes the explicit form
\begin{equation}
\kappa_\gamma^{5+10} \approx
  \frac{N_c (Q_t^2 +Q_b^2
    \hspace{0.5mm} A_{q}^h (\tau_b) )+
    A_W^h (\tau_W)+ Q_\tau^2 (
    \kappa_\tau^{5+10} \hspace{0.25mm} A_{q}^h (\tau_\tau) + 
    \hspace{0.5mm} \nu_E^{5+10}) + Q_Y^2
    \hspace{0.5mm} \nu_Y^{5+10}}{N_c (Q_t^2+Q_b^2 \hspace{0.5mm} A_{q}^h (\tau_b) ) +
    A_W^h (\tau_W)
    +Q_\tau^2 \hspace{0.5mm} A_{q}^h (\tau_\tau) }\,.
\end{equation}
The fact that, as explained before, $|\kappa_{\gamma}^{5+10}|^2$  can now become bigger than one via the potentially negative contributions due to $\nu_{Y}^{5+10}$ can be seen clearly from the plot in Figure~\ref{fig:kgam}. We discover that for $\bar \lambda \sim 20$ we can get up to a doubling in the decay cross section $h\to \gamma\gamma$.
Note however, that from the UV perspective of a GHU model, there are further constraints on the parameters. We will elaborate on this below.

\subsection{Phenomenological Implications}
\label{sec:pheno}

We finally arrive at the predictions of the models for the various Higgs channels studied at the LHC.  
In the following, we are interested in the predictions for the Higgs-production cross sections times branching ratios in the models at hand, normalized to the corresponding SM expectations
\begin{equation}
R_f^m\equiv\frac{[\sigma(pp\to h)\mathrm{Br}(h\to ff)]_{{\rm MCHM}_m}}{[\sigma(pp\to h)\mathrm{Br}(h\to ff)]_{\rm SM}}\,,
\end{equation}
for $f=\gamma,\tau,b,W,Z$, where $m=5,5$+$10$.
Moreover, we will also look at processes initiated by an explicit production mechanism of the Higgs Boson
\begin{equation}
\label{eq:Rchan}
R_f^{i;\,m}\equiv\frac{[\sigma(i)\mathrm{Br}(h\to ff)]_{{\rm MCHM}_m}}{[\sigma(i)\mathrm{Br}(h\to ff)]_{\rm SM}}\,,
\end{equation}
$i=gg\to h$, VBF, $Vh$, $tth$.
For converting the results for the Higgs decays, derived in Section \ref{sec:Higgs}, into branching fractions, one has to take into account also the change in the total decay rate $\Gamma(h)$ of the Higgs boson.
For a SM Higgs with mass $m_h\approx 125$\,GeV, the total rate is dominated by its decay into bottom quarks. Explicitly, one finds $\mbox{Br}(h\to b\bar b)\approx 0.59$, $\mbox{Br}(h\to WW)\approx 0.23$, $\mbox{Br}(h\to gg)\approx 0.07$, $\mbox{Br}(h\to \tau\tau)\approx 0.06$, and $\mbox{Br}(h\to ZZ)\approx 0.03$. In consequence, for the models at hand we arrive at
\begin{equation} 
\label{eq:RG}
   R_\Gamma^m  \equiv \frac{\Gamma(h)_{{\rm MCHM}_m}}{\Gamma(h)_{\rm SM}} \\ 
   \approx 0.59\,\big[\kappa_b^m\big]^2 
    + 0.07\,|\kappa_g^m|^2 + 0.06\,\big[\kappa_\tau^m\big]^2   + 0.28 \,.
\end{equation}
The sought ratio of the branching fractions can now be obtained as
\begin{equation}
\label{eq:BRG}
\frac{\mathrm{Br}(h\to ff)_{{\rm MCHM}_m}}{\mathrm{Br}(h\to ff)_{\rm SM}}=\frac{\Gamma(h\to ff)_{{\rm MCHM}_m}}{\Gamma(h\to ff)_{\rm SM}}/R_\Gamma^m\,.
\end{equation}
It turns out, that the changes in the decay rate for the models considered will be of minor importance since we have $\kappa_b^5=\kappa_b^{5+10}\approx\kappa_g^5=\kappa_g^{5+10}\approx 1$. We will nevertheless take them into account in the numerical analysis.

\begin{figure}[!t]
\begin{center} 
\hspace{-2mm}
\mbox{\includegraphics[height=2.75in]{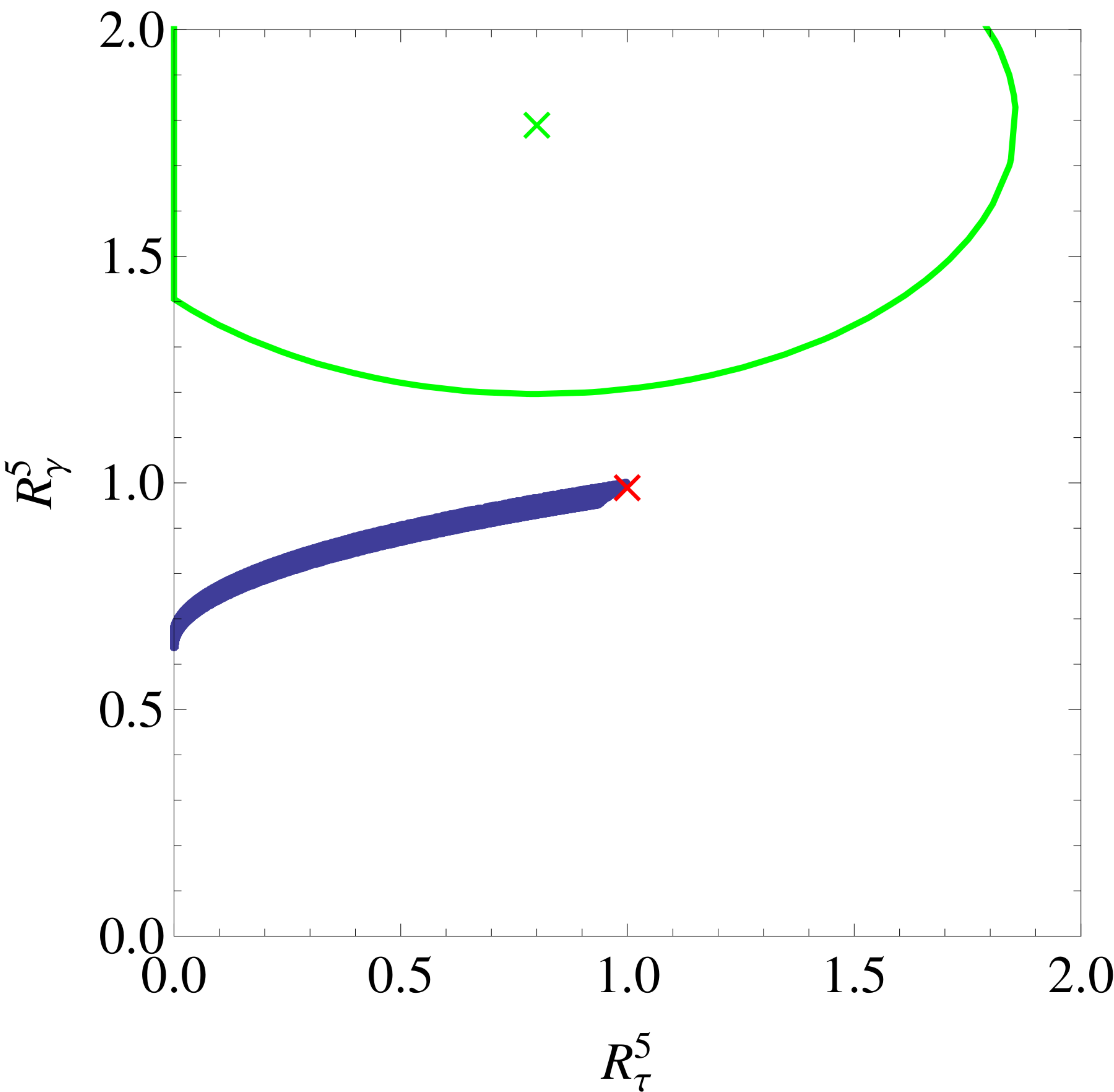}} 
\hspace{4mm}
\mbox{\includegraphics[height=2.75in]{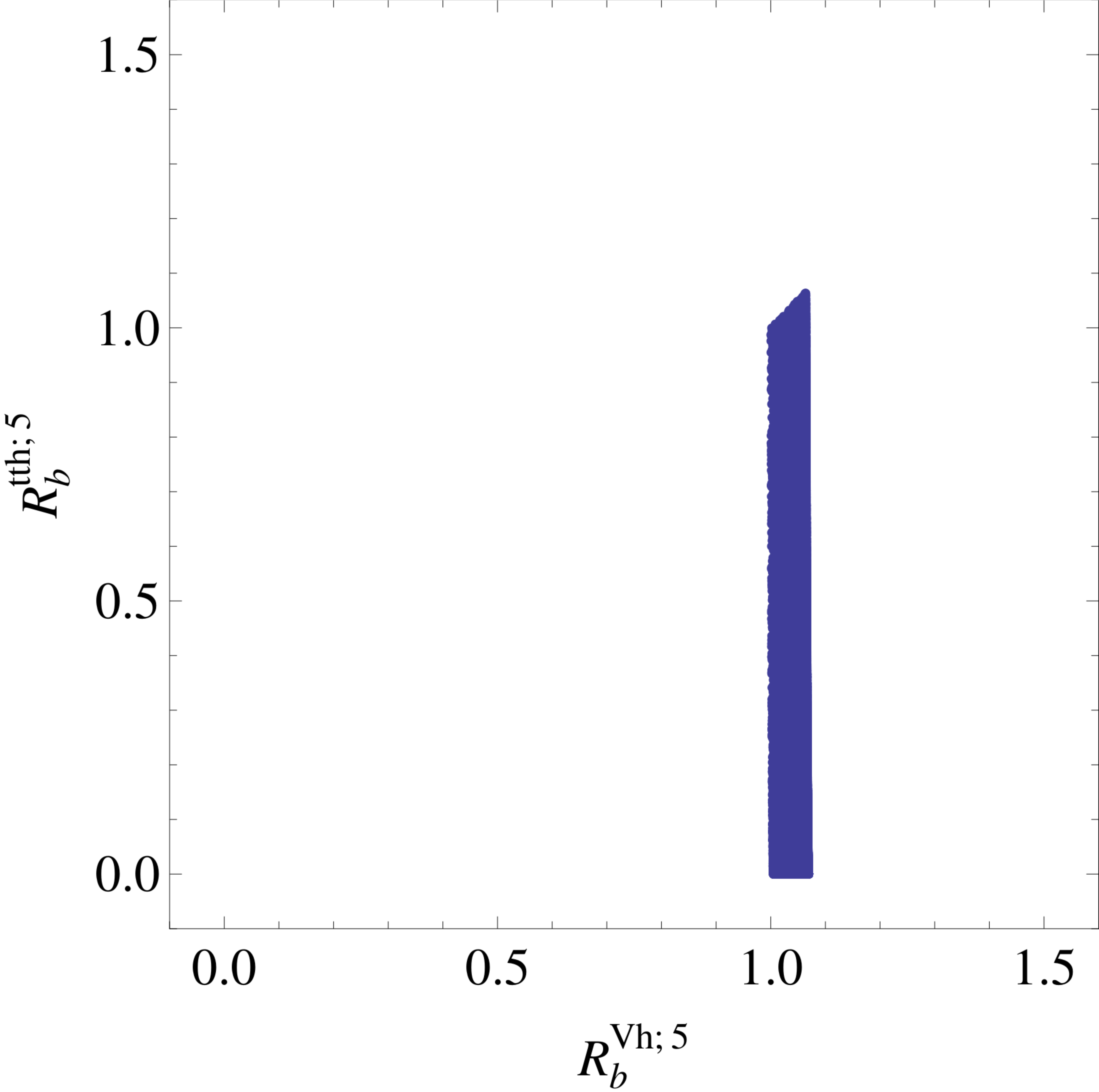}}
\parbox{15.5cm}{\caption{\label{fig:gamtaub1} Left: Prediction for the production cross section times branching fraction for $pp\to h\to \gamma\gamma$ in the MCHM$_5$ relative to the SM versus the same ratio for $pp\to h\to \tau \tau$. The experimental 1\,$\sigma$ contour from ATLAS is indicated as a green line (neglecting possible correlations). The best fit value $(R_\gamma,R_\tau)_{\rm exp}\approx(1.8,0.8)$ is shown as a green cross and the SM prediction $(R_\gamma,R_\tau)_{\rm SM}=(1,1)$ as a red cross.
Right: Higgs production cross section in the $tth$ channel times branching fraction $h\to bb$  in the MCHM$_5$ relative to the SM vs the equivalent ratio, assuming that the Higgs Boson has been produced via associated vector-boson production.}}
\end{center}
\end{figure}

We start with the results for the setup corresponding to the MCHM$_5$.
In the left panel of Figure \ref{fig:gamtaub1} we show the predictions for the change of the important discovery channel $pp\to h\to \gamma\gamma$ relative to the SM (neglecting for the moment $tth$ production)
\begin{equation}
   R_\gamma^5 \approx \frac{\big[\mbox{Br}(h\to \gamma\gamma)\big]_{{\rm MCHM}_5}}%
              {\big[\mbox{Br}(h\to \gamma\gamma)\big]_{\rm SM}} \,,
\end{equation}
versus the change in the channel $pp\to h\to \tau \tau$
\begin{equation}
   R_\tau^5 \approx \frac{\big[\mbox{Br}(h\to \tau\tau)\big]_{{\rm MCHM}_5}}%
              {\big[\mbox{Br}(h\to \tau\tau)\big]_{\rm SM}} \,.
\end{equation}
Here, we have already employed the results $\sigma({\rm VBF})_{{\rm MCHM}_m}=\sigma({\rm VBF})_{\rm SM}$ and $\sigma(V h)_{{\rm MCHM}_m}$ $=$ $\sigma(V h)_{\rm SM}$\,, see (\ref{eq:VB}) and (\ref{eq:VH}). Moreover, due to $\kappa_g^5=\kappa_g^{5+10}\approx 1$, we also could use $\sigma(gg\to h)_{{\rm MCHM}_m}\approx\sigma(gg\to h)_{\rm SM}$, see (\ref{eq:gg}).
To obtain the predictions shown in the plots, we have used the results derived from evaluating (\ref{eq:gg}), (\ref{eq:Gam}), (\ref{eq:RG}), and  (\ref{eq:BRG}). We show the full range for $0\leq c_R^{\tau, t} \leq 1$. 
For illustration, we give the experimental 1\,$\sigma$ contour from ATLAS, extracted from \cite{ATLASnote}, as a green line (neglecting possible correlations), whereas the best fit value $(R_\gamma,R_\tau)_{\rm exp}\approx(1.8,0.8)$ is shown as a green cross. The results are comparable to those of the CMS experiment. The SM prediction $(R_\gamma,R_\tau)_{\rm SM}=(1,1)$ is given as a red cross. Although it is still too early to derive too strong conclusions from the experimental situation, one can see that if the experimental errors shrink and the central values would remain in the same ballpark, the MCHM$_5$ would not be able to account for the resulting discrepancy. Going away from the SM limit ({\it i.e.} away from the decoupling limit $c_R\to 1$) leads to a stronger tension with experiment than in the SM due to the fact that in the MCHM$_5$ one can only get a reduction in the $\tau\tau$ channel together with a reduction in the $\gamma\gamma$ channel. 
The small width of the prediction for $R_\gamma^5$, for constant $R_\tau^5$, reflects the fact that the corrections due to the quark sector are of minor importance in $R_\gamma^5$, see (\ref{eq:kappagam5ap}). The strong correlation between $R_\gamma^5$ and $R_\tau^5$, both depending to good approximation only on the same parameter $c_R^\tau$ (see also below) allows to easily constrain or rule out the model, after experimental results become more precise.

\begin{figure}[!t]
\begin{center} 
\hspace{-2mm}
\mbox{\includegraphics[height=2.85in]{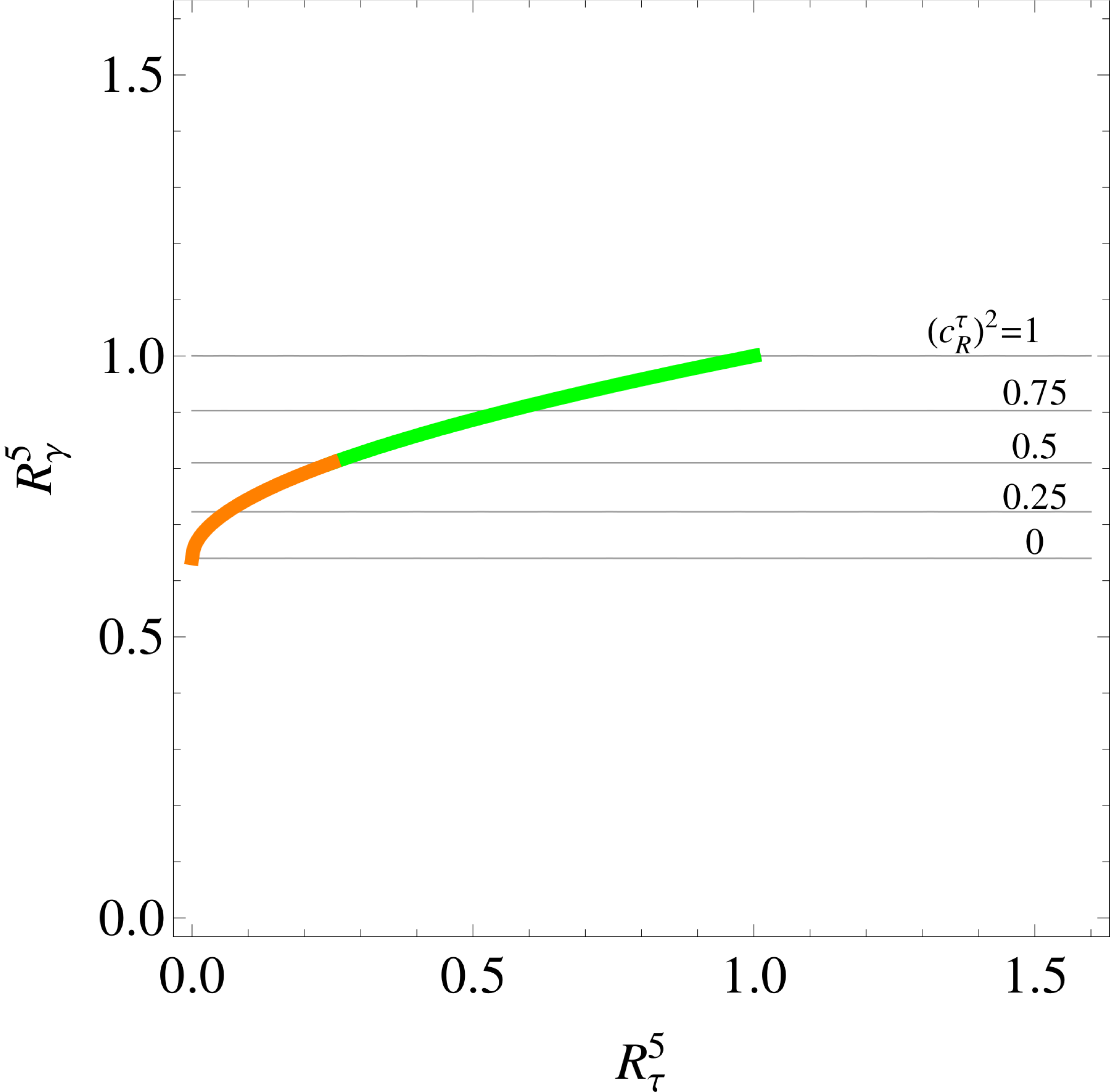}} 
\parbox{15.5cm}{\caption{\label{fig:gamtau} Left: Prediction for the production cross section times branching fraction for $pp\to h\to \gamma\gamma$ in the MCHM$_5$ relative to the SM versus the same ratio for $pp\to h\to \tau \tau$. The intersections with the horizontal lines indicate the parameter $(c_R^\tau)^2$ that results in the corresponding prediction in the $(R_\tau^5,R_\gamma^5)$-plane. See text for details.}}
\end{center}
\end{figure}

\begin{figure}[!t]
\begin{center} 
\hspace{-2mm}
\mbox{\includegraphics[height=2.85in]{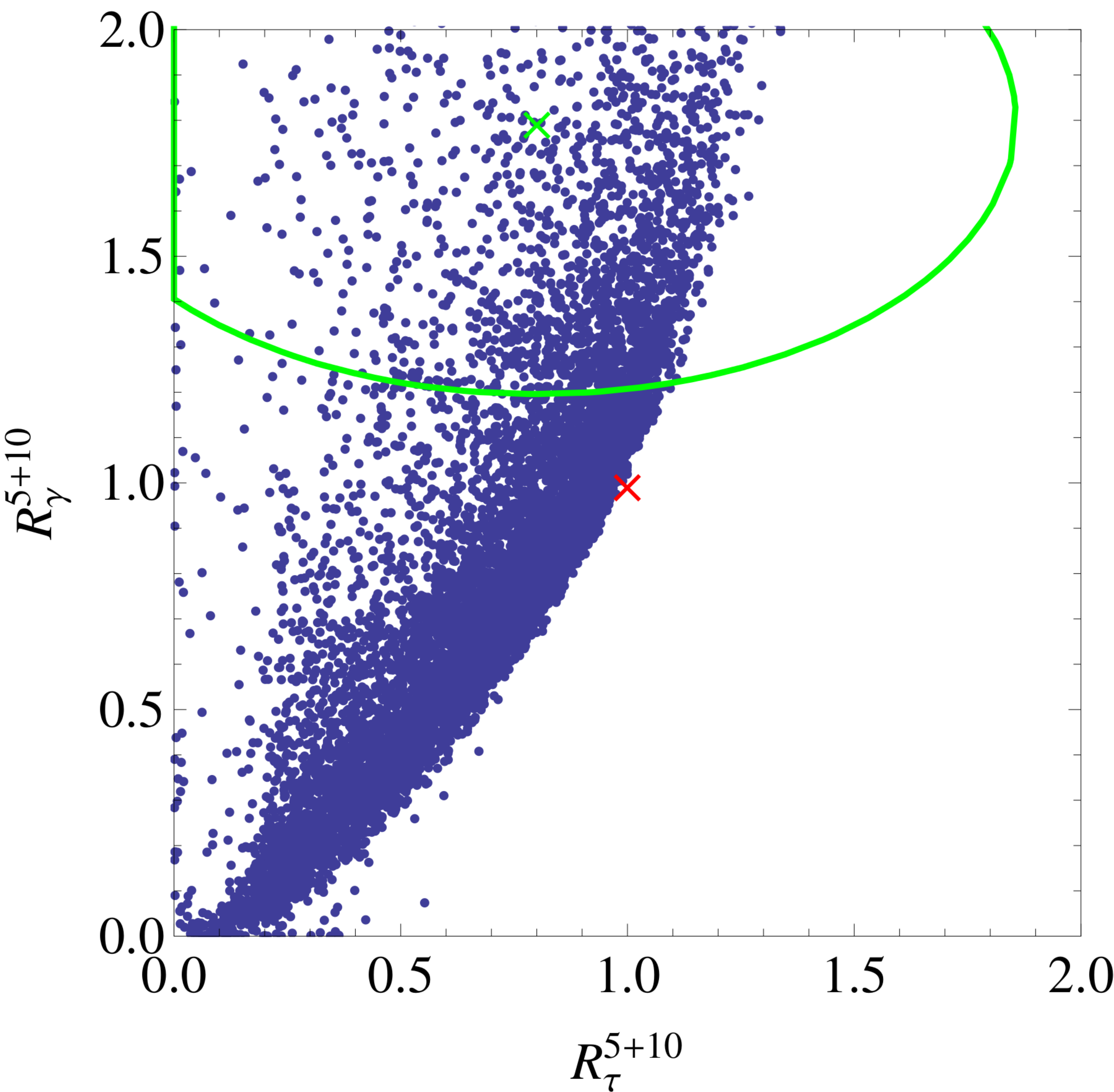}} 
\parbox{15.5cm}{\caption{\label{fig:gamtaub2} Production cross section times branching fraction for $pp\to h\to \gamma\gamma$ in the MCHM$_{5+10}$ relative to the SM versus the same ratio for $pp\to h\to \tau \tau$. The experimental 1\,$\sigma$ contour from ATLAS is again given as a green line. The best fit value $(R_\gamma,R_\tau)_{\rm exp}\approx(1.8,0.8)$ is shown as a green cross and the SM prediction $(R_\gamma,R_\tau)_{\rm SM}=(1,1)$ as a red cross. The points correspond to a scan over the parameter-space of the model.}}
\end{center}
\end{figure}

We now turn to the decay of the Higgs boson into two bottom quarks.
Note that, due to $\kappa_b^5=1$ and $\sigma(V h)_{{\rm MCHM}_m}=\sigma(V h)_{\rm SM}$,  the process $q\bar q^{(\prime)}\to V^\ast \to V h$, $V=W,Z$, with a subsequent decay $h\to b \bar b$ remains SM-like to good approximation (the deviation of $R_\Gamma^5$ from 1 due to $\kappa_\tau^5\neq1$ is only at the level of a few per cent)
\begin{equation}
R_b^{Vh;\, 5}\approx 1.
\end{equation}
However, due to $\kappa_t^5<1$ the search channel  $gg \to t \bar t^\ast t^\ast \bar t \to t \bar t h$, $h\to b \bar b$ can receive a sizable suppression, which is illustrated in the right panel of Figure \ref{fig:gamtaub1},
where we show  $R_b^{tth;\, 5}$ versus $R_b^{Vh;\, 5}$. Once the experimental situation in these channels improves they will become a superb tool for measuring directly a possible reduction in the $\bar t t h$ coupling. This prediction of the MCHM$_5$, together with an expected more SM-like behavior in the $Vh$ channel (both in tentative agreement with latest CMS results \cite{CMSnote}), would allow for another possibility to test the model. 
Since the formulas in the MCHM$_5$ are easy enough, it is even possible to solve for the important parameters of the model, $c_R^t$ and $c_R^\tau$, in dependence on the $R_f^{(i;)\,5}$. 
Using (\ref{eq:tth}) and (\ref{eq:Rchan}) we directly obtain the approximate relation
\begin{equation}
(c_R^t)^2\approx\sqrt{R_b^{tth;\,5}}\,.
\end{equation}
In the same way one can solve for $c_R^\tau$, using the information from $R_\gamma^5$ and $R_\tau^5$. The approximate result, neglecting corrections due to $A_{q}^h (\tau_t)\neq 1$ $A_{q}^h (\tau_\tau)\neq 0$ and $R_\Gamma^5\neq 1$, is shown in Figure \ref{fig:gamtau} for illustration.
The intersection of the straight lines, corresponding to different $c_R^\tau$, with the prediction in the $(R_\tau^5,R_\gamma^5)$-plane, depicted by the colored line, gives the MCHM$_5$ result in dependence on the input parameter $c_R^\tau$. This would allow to get insight about the possible parameters of the extended fermion sector, once a more precise measurement (compatible with the predictions of the model) would be established. The orange color indicates values of $c_R^\tau<1/\sqrt 2$ that correspond to a large compositeness of the order of $m^\prime \sim M$ and should be taken with caution.

Finally, it is clear from the previous discussion that also, if the Higgs boson is produced in gluon-gluon fusion, VBF or associated $W^\pm$-production, the double-vector boson production through a Higgs remains unchanged to good approximation in the MCHM$_5$, see (\ref{eq:WZ}), which is in reasonable agreement with first measurements of the experiments \cite{ATLASnote,CMSnote}.

\begin{figure}[!t]
\begin{center} 
\hspace{-2mm}
\mbox{\includegraphics[height=2.85in]{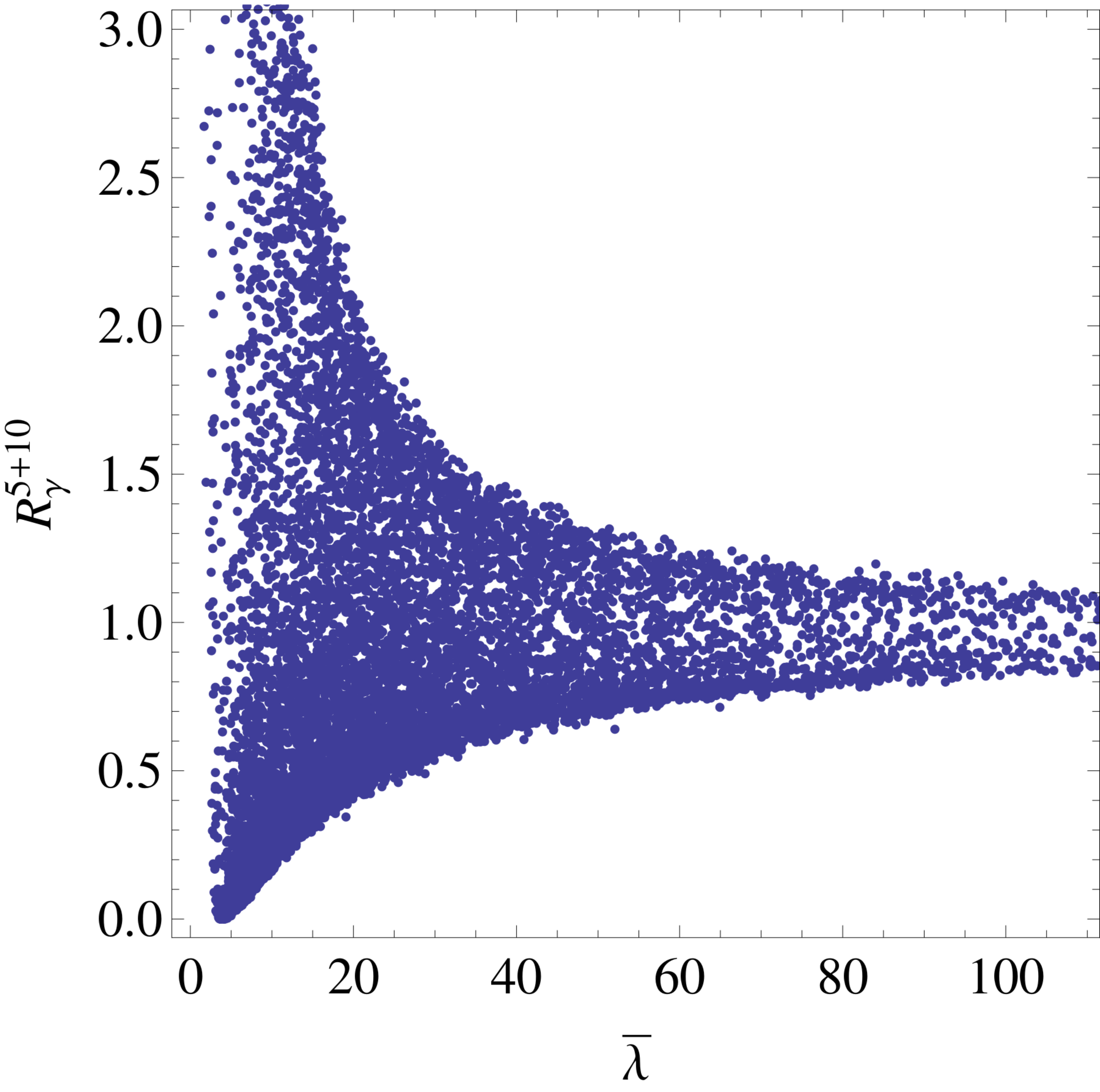}} 
\quad
\mbox{\includegraphics[height=2.85in]{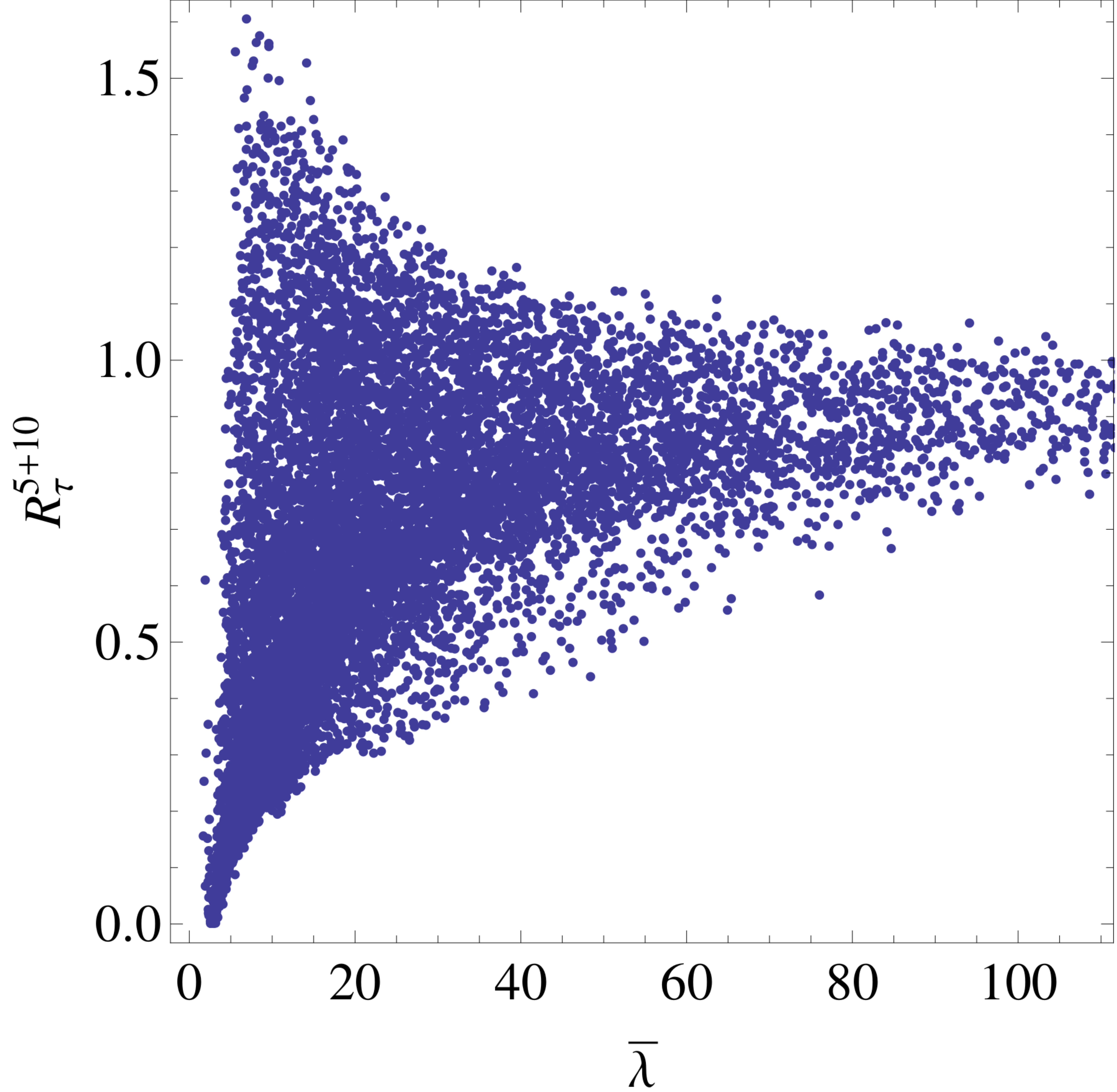}}
\parbox{15.5cm}{\caption{\label{fig:gamtau3}
Left: Predictions for $R_\gamma^{5+10}$  plotted versus the parameter $\bar \lambda=2 M \tilde M/(v^2 |y^\prime \bar y|)$, measuring the vector-like masses in the lepton sector over the product of the compositeness of the right handed $\tau$ with the Yukawa couplings of the new leptonic resonances. Right: Analogous plot, now for $R_\tau^{5+10}$. The points correspond to a scan over the parameter-space of the model. See text for details.}}
\end{center}
\end{figure}

We now move to the discussion of the MCHM$_{5+10}$. As explained in Section \ref{sec:Higgs10}, we expect different predictions for this version of the model.
We start again by studying the correlation between $R_\gamma^{5+10}$ and $R_\tau^{5+10}$, which is depicted in the plot in Figure \ref{fig:gamtaub2}. Now, as detailed in Section \ref{sec:Higgs10}, the decay into photons can receive an enhancement with respect to the SM. Moreover, also in the MCHM$_{5+10}$ one gets the rough prediction $R_\tau^{5+10}<1$. Taken together this allows, in contrast to the MCHM$_5$, for the possibility to reach the best fit value $(R_\gamma,R_\tau)_{\rm exp}\approx(1.8,0.8)$  in the MCHM$_{5+10}$. Note that this is possible without spoiling the rough agreement with the SM in the other channels, which are (besides $tth$ production) unchanged to good approximation in the models at hand.  Moreover, although due to the new $Q=-2$ resonances the correlations are not as strong as in the MCHM$_5$, still, finding e.g. a reduced $\gamma\gamma$ signal, together with an enhanced $\tau\tau$ rate, would exclude the model. To judge which scales for the parameters of the model lead to which prediction, we give in the left plot of Figure \ref{fig:gamtau3} our result for $R_\gamma^{5+10}$ versus the parameter $\bar \lambda$ (see (\ref{eq:lamb})), measuring the vector-like masses over the product of the compositeness of the right handed $\tau$ with the Yukawa couplings of the new leptonic resonances. This parameter is still appropriate, as the impact of the new quark sector on $R_\gamma^{5+10}$ is very limited. We deduce that for moderately low scales $\bar \lambda \sim 20$, an enhancement of up to a factor of 2 in the $pp\to h \to \gamma\gamma$ channel is possible (see \cite{ArkaniHamed:2012kq} in this context). We stress that, although significant changes in the couplings of the leptons to the Higgs sector appear, the agreement with electroweak precision observables is saved due to the custodial symmetry. In the right panel of the figure, we show the analogous plot for $R_\tau^{5+10}$. We observe that a reduction $R_\tau^{5+10}<0.5$ would correspond to scales $\bar \lambda <20$ (a non-negligible enhancement possible for such low scales comes with extremely large values of  $R_\gamma^{5+10}$). Such low scales would still be viable, given that the vector-like masses themselves are not beyond the TeV scale.
Note that also in the MCHM$_{5+10}$ the mass eigenvalues of the resonances alone have only a limited impact on the size of the effects, as a larger mass can be compensated by larger Yukawa couplings. The question then becomes how large one could assume the values for the Yukawa couplings. For $\bar \lambda$ of ${\cal O}(1)$, these Yukawas would be of the order of the masses of the heavy resonances over $v$ which could become problematic.
Finally, concerning the decay into bottom quarks, the predictions in the MCHM$_{5+10}$ are comparable to those in the in the MCHM$_{5}$, due to the unchanged quark sector.

\begin{figure}[!t]
\begin{center} 
\hspace{-2mm}
\mbox{\includegraphics[height=2.85in]{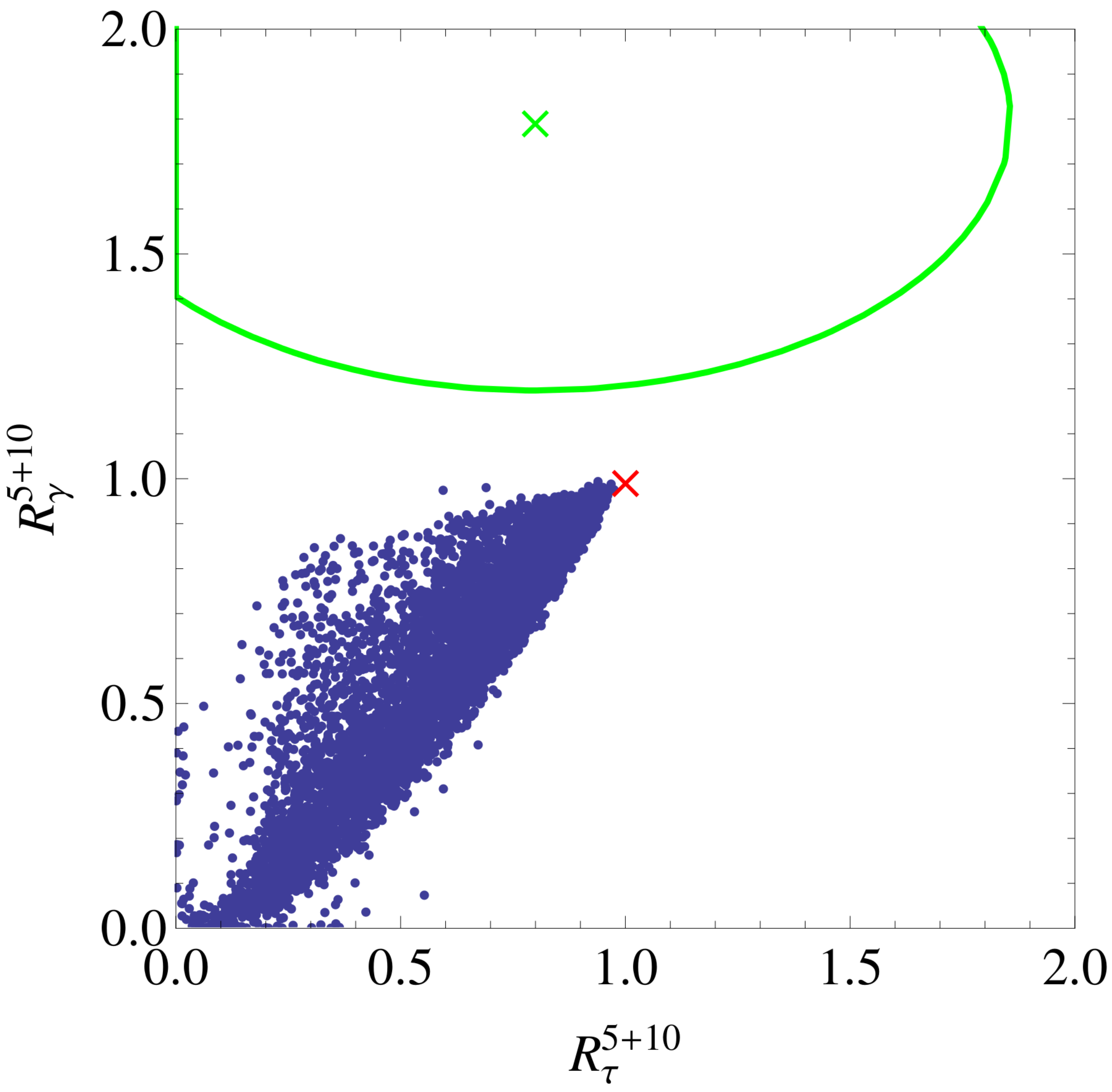}} 
\parbox{15.5cm}{\caption{\label{fig:gamtaub2cut} Production cross section times branching fraction for $pp\to h\to \gamma\gamma$ in the MCHM$_{5+10}$ relative to the SM versus the same ratio for $pp\to h\to \tau \tau$. The experimental 1\,$\sigma$ contour from ATLAS is again given as a green line. The best fit value $(R_\gamma,R_\tau)_{\rm exp}\approx(1.8,0.8)$ is shown as a green cross and the SM prediction $(R_\gamma,R_\tau)_{\rm SM}=(1,1)$ as a red cross. The points correspond to a scan over the parameter-space of the model with the constraint $\nu_Y^{5+10}>0$ from GHU models.}}
\end{center}
\end{figure}

We have seen that in the MCHM$_{5+10}$ the strong correlations present in the MCHM$_5$ are washed out, making on the one-hand side the model less predictive but on the other allow in principle for a better agreement with preliminary results from the LHC. Still, as discussed above, correlations and empty regions of parameter space remain, allowing to test the model. 

\begin{figure}[!t]
\begin{center} 
\hspace{-2mm}
\mbox{\includegraphics[height=2.82in]{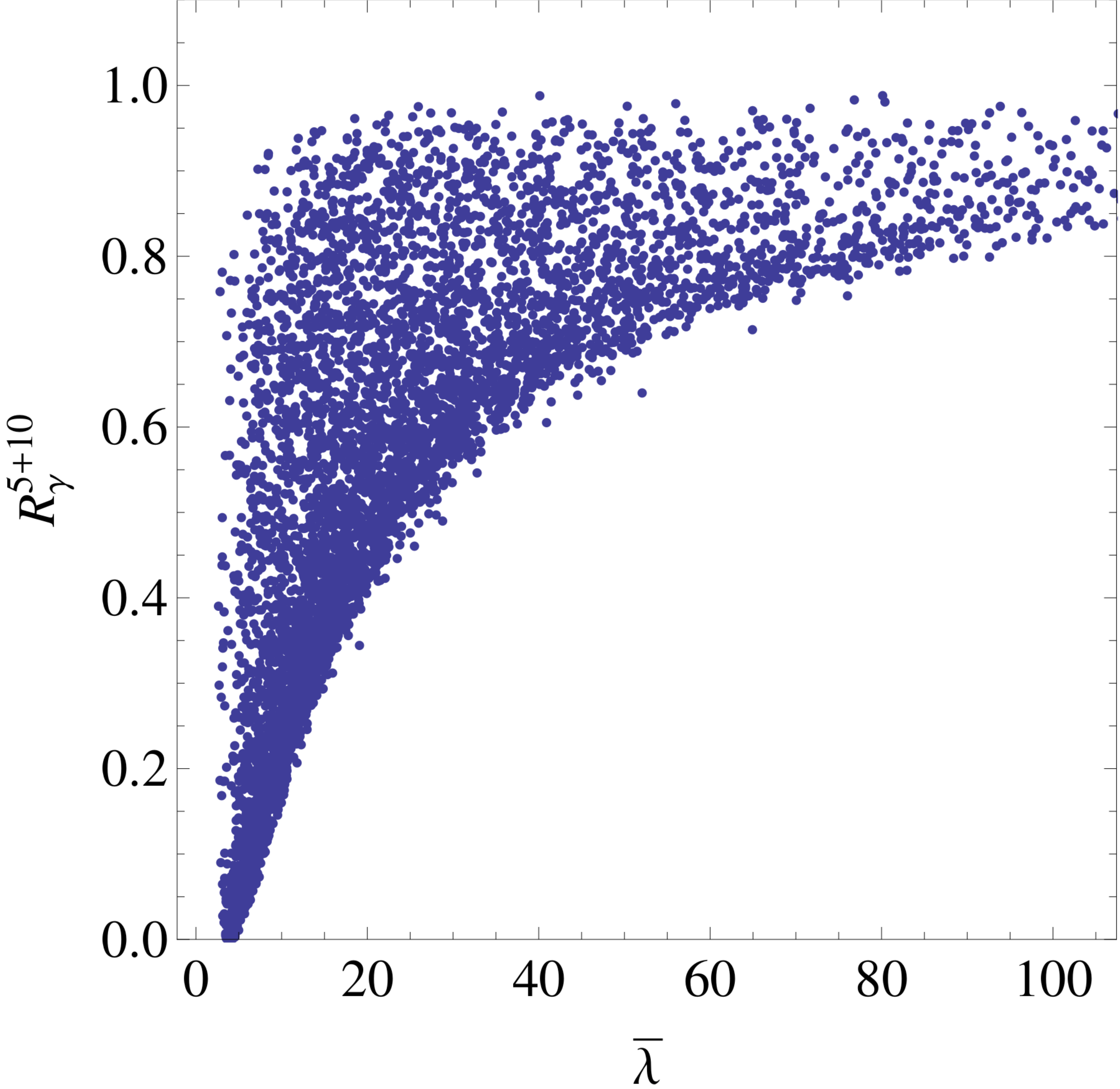}} 
\quad
\mbox{\includegraphics[height=2.82in]{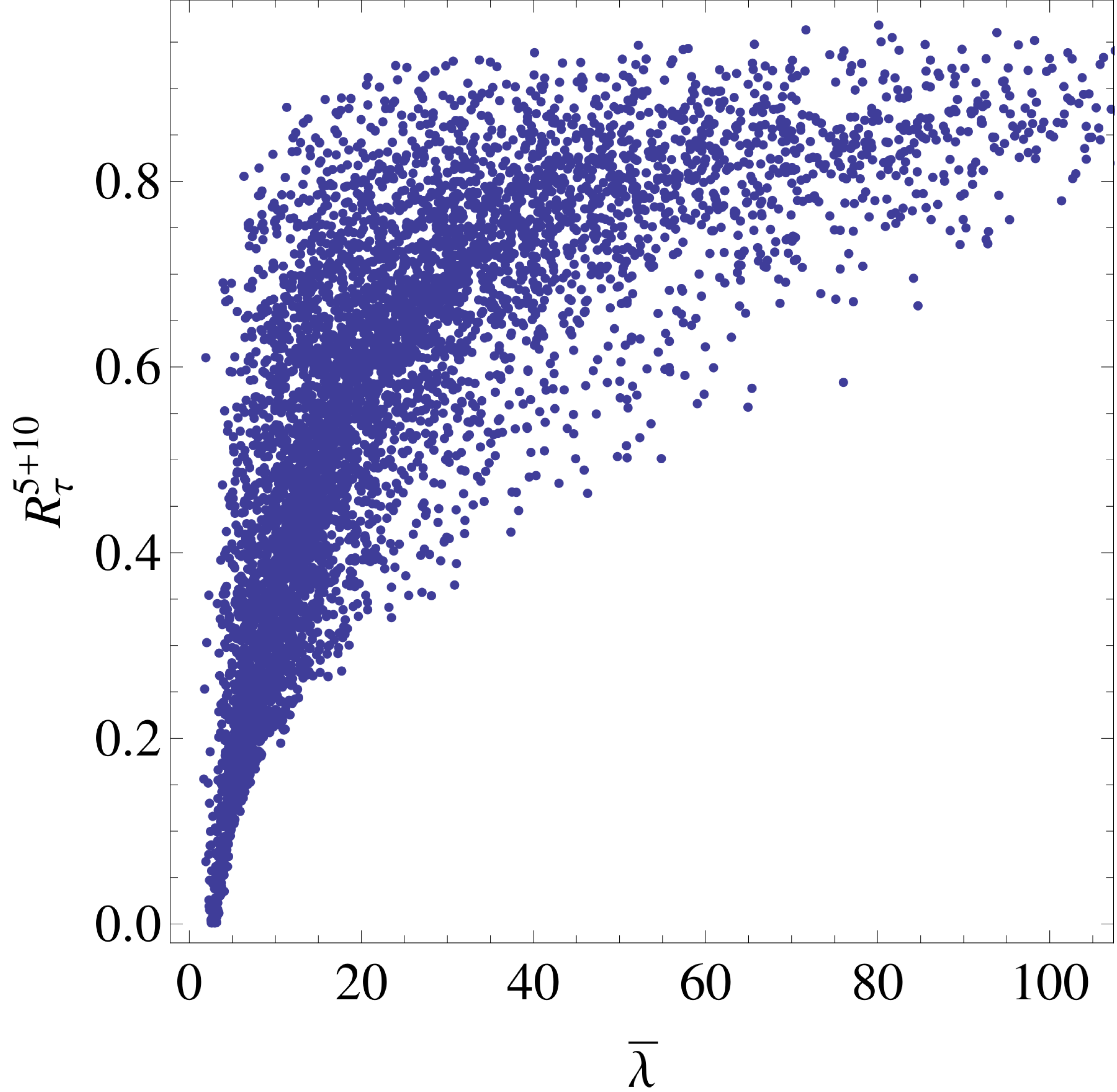}}
\parbox{15.5cm}{\caption{\label{fig:gamtau3cut}
Left: Predictions for $R_\gamma^{5+10}$  plotted versus the parameter $\bar \lambda$.
Right: Analogous plot, now for $R_\tau^{5+10}$. The points correspond to a scan over the parameter-space of the model, employing the constraint $\nu_Y^{5+10}>0$ from GHU models. See text for details.}}
\end{center}
\end{figure}

So far, we have constrained the parameters of the model roughly from naturalness arguments and phenomenology. However, as already mentioned before, in GHU models there are more correlations present between the parameters. For example such correlations will lead to relations between the numerator and the denominator of $\bar \lambda$, making it not a completely free parameter anymore, and the same holds true for $c_R^{t,\tau}$. The results given above hence are more general in comparison to considering GHU models as a UV completion \cite{next}. The predictions include those of GHU models as a subset. Moreover, it turns out that always $\nu_Y^{5+10}>0$ in GHU, making the model more predictive with respect to $R_\gamma^{5+10}$.
In Figures \ref{fig:gamtaub2cut} and \ref{fig:gamtau3cut} we plot again the same quantities given before, now employing this condition.

First, we can see that the condition eliminates the small amount of points corresponding to 
$R_\tau^{5+10}>1$. This can be also understood from (\ref{eq:0mode}) and (\ref{eq:KKsum}). Implementing the information from the GHU model leads to ${\rm Re}(\bar y \hat y) \approx - {\rm Re}(\bar y y^\prime) >0$ (see (\ref{det:MYA})) and thus, directly from  (\ref{eq:0mode}) we get $\kappa_\tau^{5+10}<1$. Moreover, all the parameter-space with $R_\gamma^{5+10}>1$ is gone, as expected. In Figure \ref{fig:gamtau3cut} we give the dependence of the individual quantities on $\bar \lambda$ in the constrained setup for completeness. Seeing our MCHM$_{5+10}$ as the low energy limit of a GHU model thus leads again to the robust prediction of $R_{\gamma}^{5+10},R_{\tau}^{5+10}<1$. Exploring the $(R_\gamma,R_\tau)$-plane experimentally can on the one hand give hints if an extended fermion sector featuring custodial protection, as studied in this work, could exist and on the other hand could also say something about the possible UV completion of such a sector.

Finally, as in the MCHM$_{5}$, due to the SM-like coupling of the Higgs boson to gauge bosons (and bottom quarks), the double-vector boson production through a Higgs also remains unchanged to good approximation if the Higgs boson is produced in gluon-gluon fusion, VBF or associated vector-boson production.

\subsection{Impact of the Non-Linearity of the Higgs}

\label{sec:vf}

Up to now, we have neglected the effects arising from the 
pseudo-Goldstone boson nature of the Higgs boson in the corresponding UV 
completions of our models. Considering this will lead to shifted Higgs couplings to 
the different fermions and gauge bosons of the spectrum. In the latter 
case, neglecting the mixing of the SM gauge bosons to their composite 
counterparts, everything is fixed by the quantum numbers and the 
symmetry breaking defining the composite model. For the models that we 
study we obtain \cite{Giudice:2007fh}
\begin{eqnarray}
\kappa_W^{m}=\kappa_Z^{m}=\cos\left(\frac{v}{f}\right)\approx 
\sqrt{1-\xi},\qquad m=5,5+10,
\end{eqnarray}
where we have defined $\xi=v^2/f^2$ as usual.

In the case of fermions, we have to make the difference between the two 
cases considered in this paper, since the explicit expressions for these 
additional corrections depend also on the different representations 
chosen for fermions. In the MCHM$_5$, the corresponding modifications of 
the Higgs couplings to SM fermions read
\begin{eqnarray}
\kappa_f^5\to 
\kappa_f^5\cos\left(\frac{2v}{f}\right)/\cos\left(\frac{v}{f}\right)\approx 
\kappa_f^5(1-2\xi)/\sqrt{1-\xi},
\end{eqnarray}
where $f$ is running over all the fermions of the SM. %Similarly,
%\begin{eqnarray}
%\nu_F^5\to \nu_F^5\cos\left(\frac{2v}{f}\right)/\cos\left(\frac{v}{f}\right)\approx \nu_F^5(1-2\xi)/\sqrt{1-\xi},
%\end{eqnarray}
%with $F=T,E$.
Here and in the following we neglect $v^2/f^2$ corrections to 
contributions which are already suppressed by the new physics scale. 
In this approximation, all the $\nu_F^m$ factors remain unchanged. The 
previous shifts in the fermion couplings to the Higgs boson will lead 
also to a suppression of the effective coupling to gluons, which can be written as
\begin{eqnarray}
\kappa_{g}^5\approx\cos\left(\frac{2v}{f}\right)/\cos\left(\frac{v}{f}\right) \approx (1-2\xi)/\sqrt{1-\xi}.
\end{eqnarray}
In the case of the MCHM$_{5+10}$ the expressions have to take into 
account that the $\tau_R$ is living in a $\mathbf{10}$ while the opposite
chirality is embedded in a fundamental representation of $SO(5)$. This 
leads to \cite{Azatov:2011qy}
\begin{eqnarray}
\label{eq:ktauc}
\kappa_{\tau}^{5+10}\to 
\kappa_{\tau}^{5+10}\cos\left(\frac{v}{f}\right)\approx 
\kappa_{\tau}^{5+10}\sqrt{1-\xi}
\end{eqnarray}
%and
%\begin{eqnarray}
% \nu_{E}^{5+10}\to \nu_{E}^{5+10}\cos\left(\frac{v}{f}\right)\approx \nu_E^{5+10}\sqrt{1-\xi},
%\end{eqnarray}
while the other couplings change analogously as in the MCHM$_5$.
Finally, the change in $\kappa_\gamma^m$ can be worked out for both models by applying the replacements given above to 
(\ref{eq:kappagamma}) (including the change in the Higgs coupling to $W^\pm$-bosons).

\begin{figure}[!t]
\begin{center} 
\hspace{-2mm}
\mbox{\includegraphics[height=2.75in]{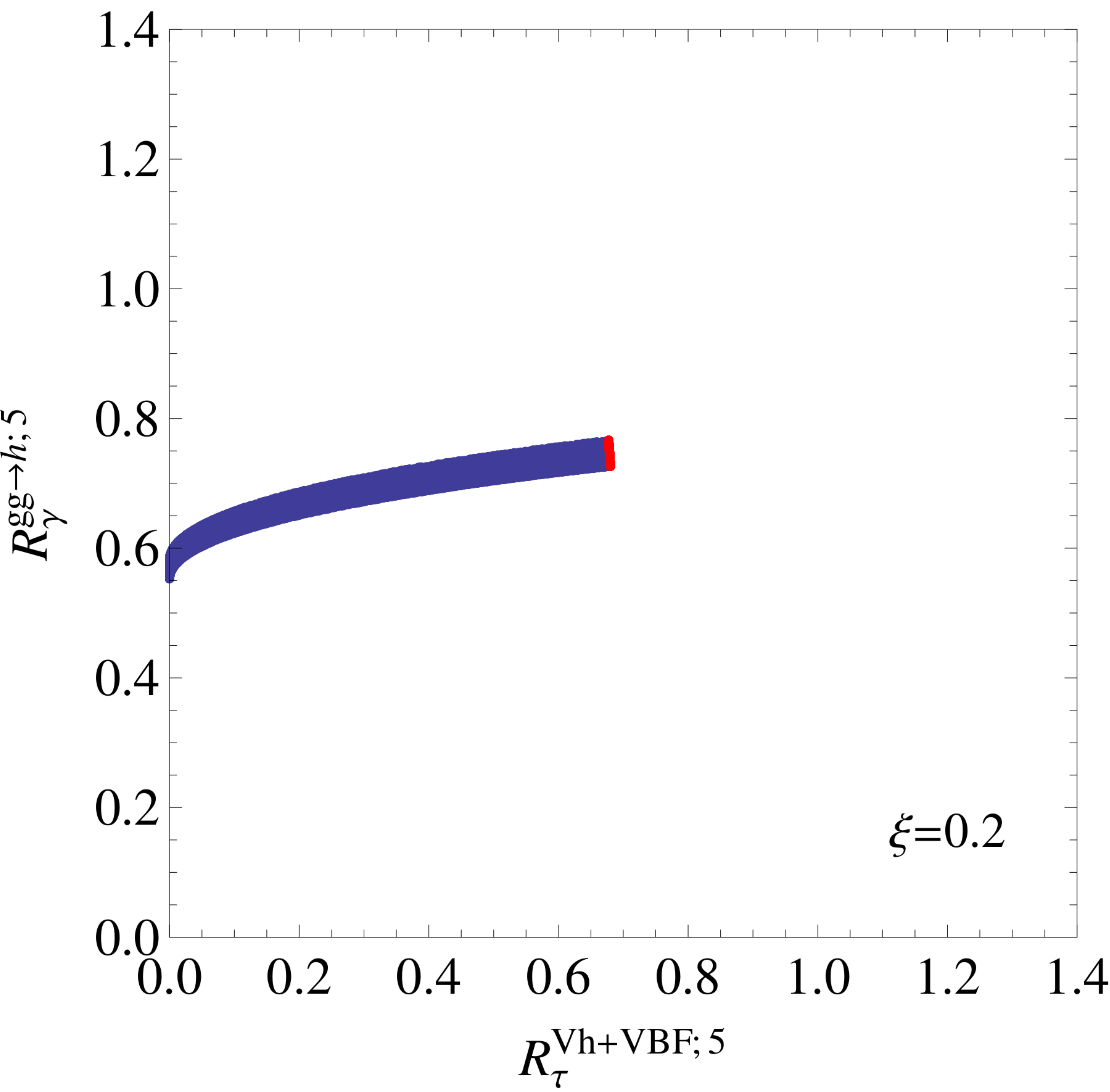}} 
\hspace{4mm}
\mbox{\includegraphics[height=2.75in]{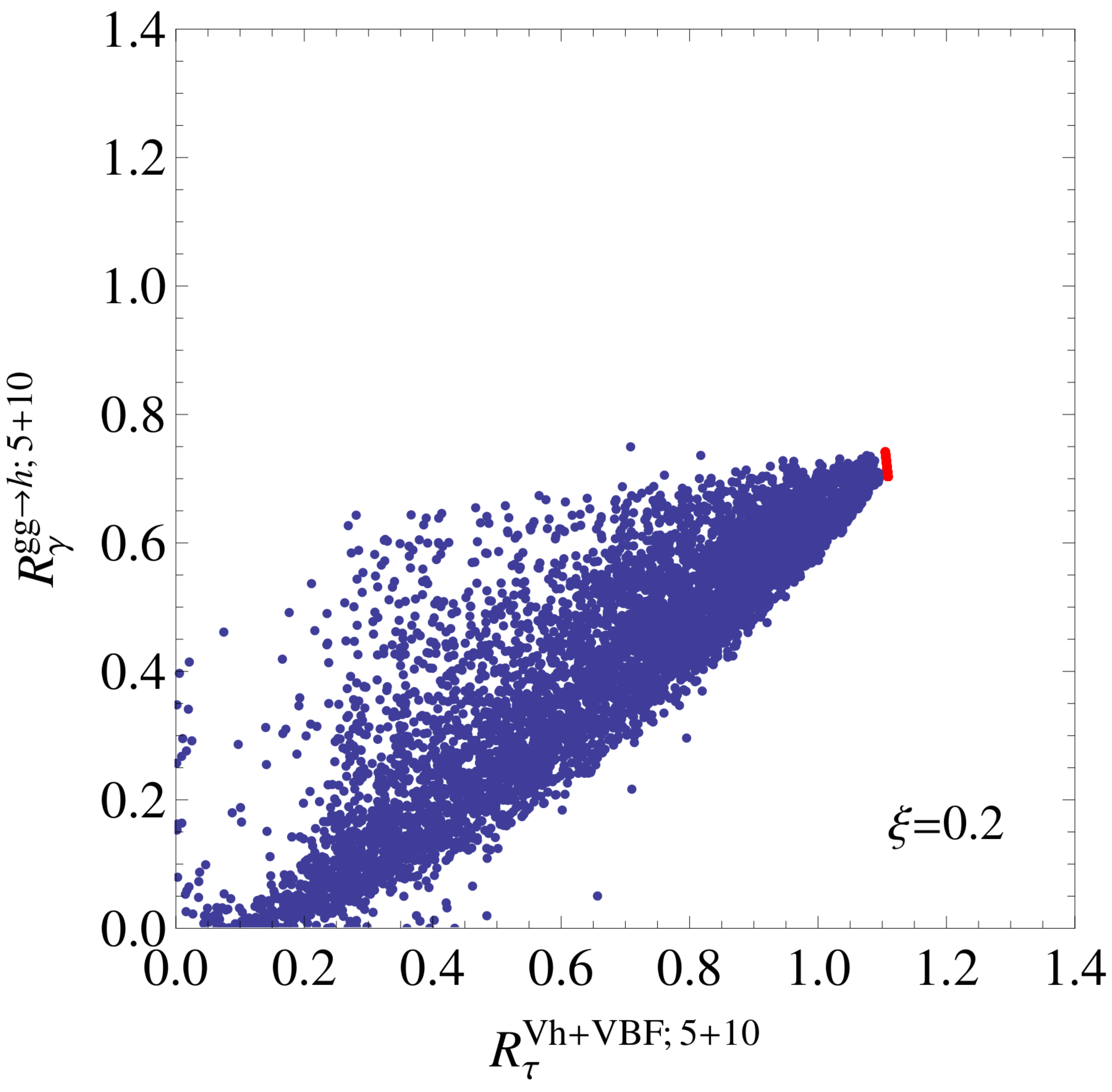}}
\parbox{15.5cm}{\caption{\label{fig:gamtaun} Left: Prediction for the production cross section times branching fraction for $gg$ $\to$ $ h\to \gamma\gamma$ in the MCHM$_5$ including leading order effects from the non-linearity of the Higgs sector, relative to the SM versus the same ratio for $h\to \tau \tau$ in VBF or $Vh$ production. The red region corresponds to the prediciton neglecting the mixing with the composite lepton sector.
Right: The analogous plot for the MCHM$_{5+10}$.}}
\end{center}
\end{figure}

\begin{figure}[!t]
\begin{center} 
\hspace{-2mm}
\mbox{\includegraphics[height=2.75in]{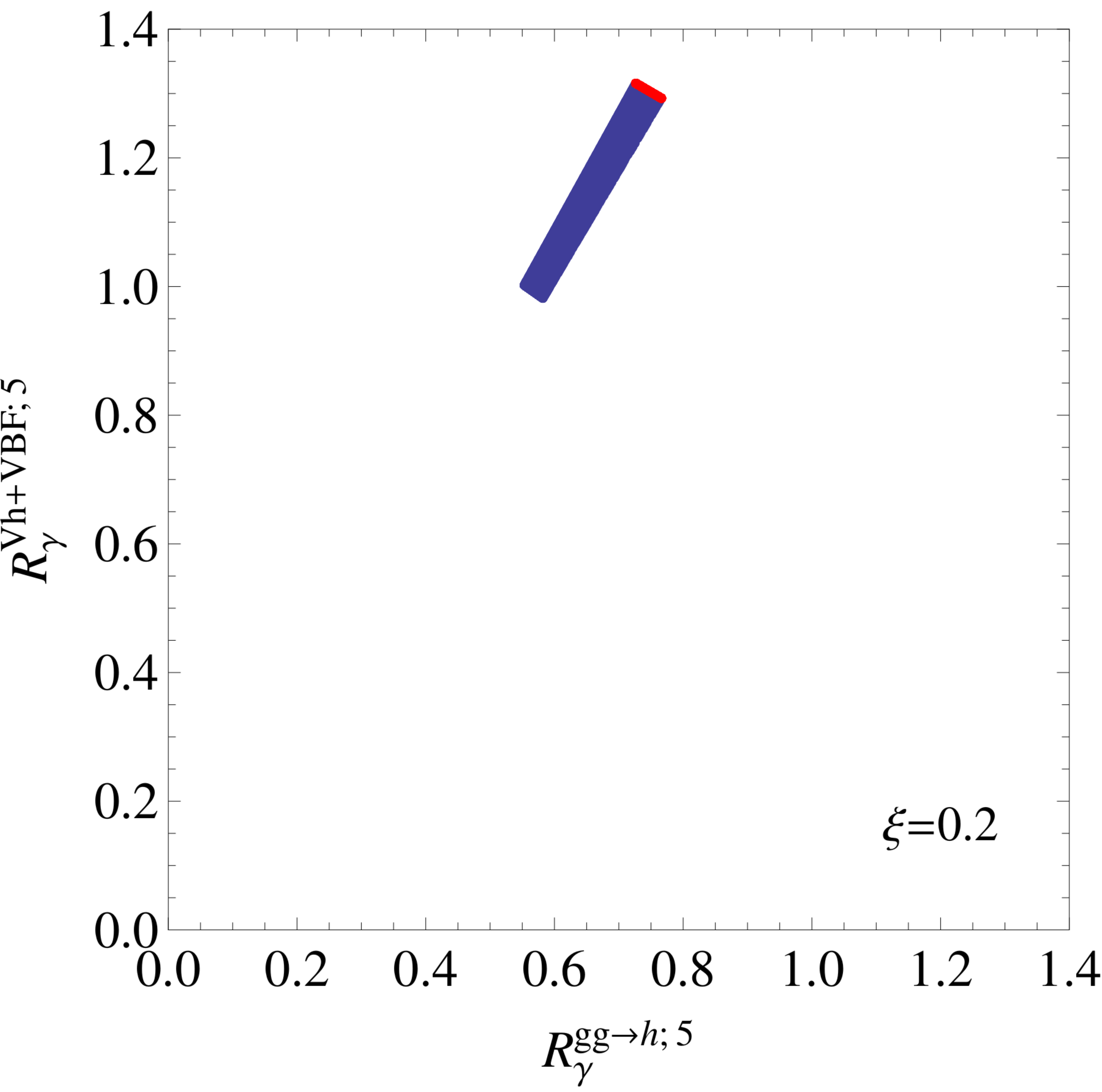}} 
\hspace{4mm}
\mbox{\includegraphics[height=2.75in]{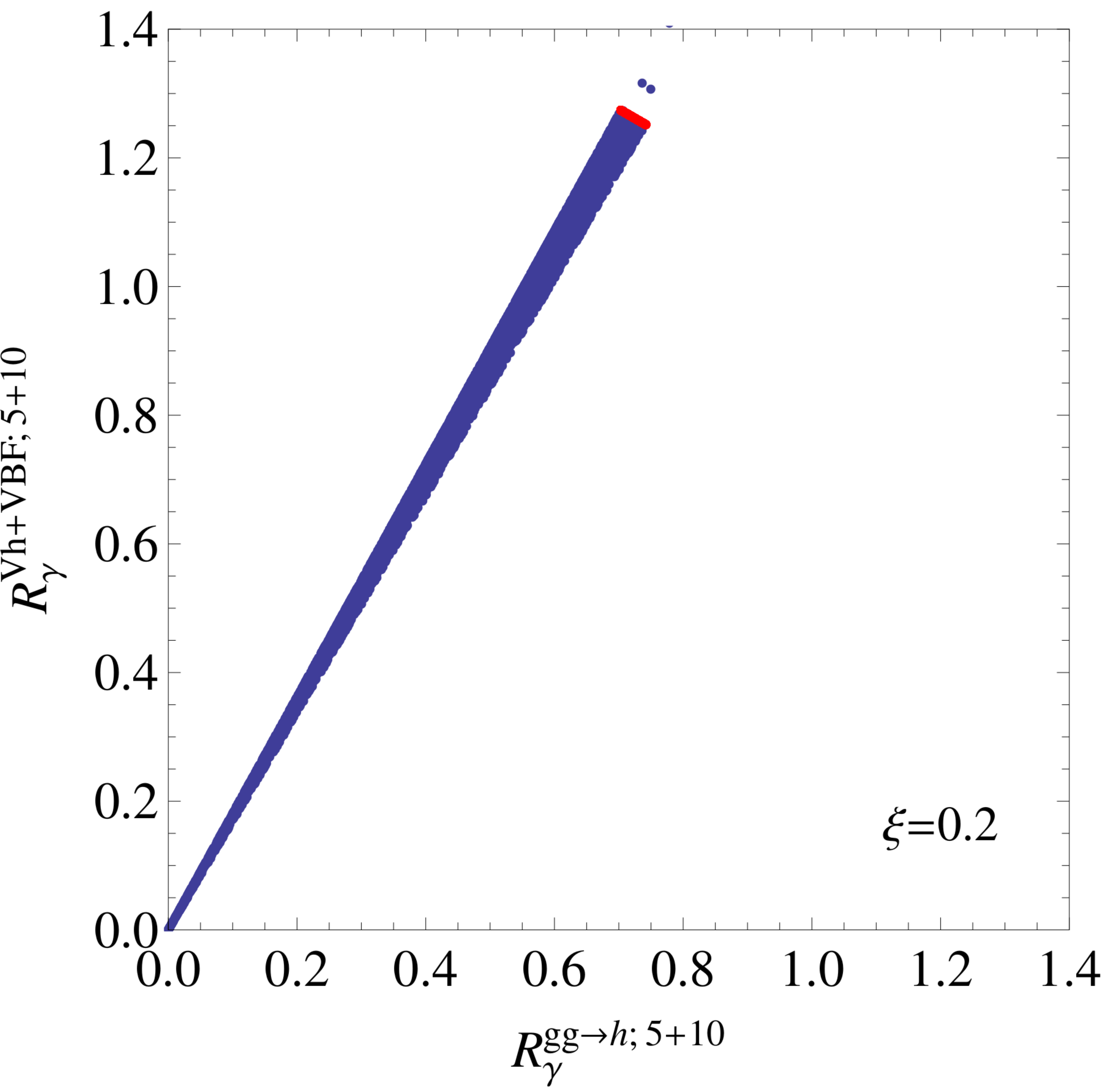}}
\parbox{15.5cm}{\caption{\label{fig:V} Left: Prediction for the production cross section times branching fraction for $gg$ $\to$ $ h\to \gamma\gamma$ in the MCHM$_5$ with respect to the same decay if the Higgs has been produced in VBF or $Vh$ production,
including leading order effects from the non-linearity of the Higgs sector. The red region corresponds to the prediciton neglecting the mixing with the composite lepton sector.
Right: The analogous plot for the MCHM$_{5+10}$.}}
\end{center}
\end{figure}

We have implemented all these additional corrections in our 
phenomenological study, employing $\xi=0.2$, to see to what extend the previous picture is 
changed. The neglected effects
arising from the non-linearity of the Higgs are a subleading correction 
to the shift in the Yukawa couplings for both the
top quark and the $\tau$ lepton in the regime where both
fermions are mostly composite and strongly interact with the different
vector-like resonances, although it can be important in the decoupling 
limit.\footnote{We should stress that even for the more predictive and 
constrained 5D picture, we can still make the coupling of the SM 
fermions to the composite sector small without the need to reduce $\xi$. 
For instance, this can be achieved by UV localizing the corresponding 
fermions and thus making the interaction with the KK modes small.}
Comparing the plot in the left panel of  Figure \ref{fig:gamtaun} to the equivalent one 
shown before, see Figure \ref{fig:gamtaub1}, we can see that taking into account the pseudo-Goldstone character of
the Higgs boson leads to a further reduction in both the $gg \to h \to \gamma \gamma$
and $\tau\tau$ channels in the MCHM$_5$, where for the latter the Higgs is assumed to be produced in VBF or $Vh$ production. Even though this is not shown in the figure, the same holds for gluon-gluon fusion. The different trigonometric rescaling appearing for the $\tau$ in the MCHM$_{5+10}$, see (\ref{eq:ktauc}), leads to modifications to this picture, as can be seen comparing the right panel of the previous figure to Figure \ref{fig:gamtaub2cut}. In this case, the larger suppression for the other fermions can in principle enhance $h\to \tau \tau$ through VBF and $Vh$ production. This is not longer true if we consider other production mechanisms like $gg\to h$ or $tth$. A similar enhancement can happen in both models for $h\to \gamma \gamma$ in the production mechanisms induced by weak gauge bosons, as is shown in Figure \ref{fig:V}, because the smaller trigonometric suppressions of the couplings of the latter still allow for an enhancement in the production cross section times branching fraction, despite the reduction in $\kappa_\gamma^m$ due to the composite fermions.
We should notice however that these plots are made using a moderately large value of $\xi=0.2$, which has to be compared for instance with the one arising from a 5D construction with a small KK scale of 1.5\,TeV, which is $\xi\approx 0.1$. Concerning the changes in Higgs decays to bottom quarks, the trigonometric rescalings of the fermion interactions lead to a reduction in both $R_b^{Vh;\,5}$ and the maximum possible $R_b^{tth;\,5}$.
Finally, as can be seen from the red regions in the last two figures, which show the predictions neglecting the impact of light lepton custodians, their effect is important and should be taken into account.

\section{Conclusions}
\label{sec:conclusions}

We have studied the impact of modified fermion sectors, featuring a custodial symmetry that protects the $Zb_Lb_L$ and $Z \tau_R \tau_R$ vertices, on Higgs production and decay. On the one hand, these setups can be thought of as simple extensions of the SM, viable on their own. On the other hand, they can be particularly motivated as the low energy tail of composite Higgs models (MCHM$_{5,10}$) or models of gauge-Higgs unification. Here the particles we consider arise as light custodians, associated to the significantly composite top quark and $\tau$ lepton.
Due to the simple structure of the setups considered, we were able to clearly relate our predictions to the model parameters. Moreover, as we explained, this framework allows to capture the physics of possible UV completions, e.g. models of gauge-Higgs unification, in a simplified way. Due to the full consideration of a realistic (composite) lepton sector for the first time in the context of Higgs signals, we found a distinct phenomenology with respect to previous studies of composite models. In particular, we discovered generically a large reduction for the Higgs decay into two $\tau$ leptons in both setups considered, which is interesting in the light of the fact that a reduced $\tau$-signal still fits well with the data \cite{ATLASnote,CMSnote}. 
On the other hand, neglecting possible UV completions, the new leptons of our framework of embedding the $\tau_R$ in a \textbf{10} of $SO(5)$ allow in principle for an enhancement in the channel $pp\to h\to \gamma\gamma$, in agreement with current results from the LHC \cite{ATLASnote,CMSnote}.
If the experimental trend is confirmed, it would favor such an embedding with respect to the one of the MCHM$_5$.

However, considering the setups as the low energy theories of gauge-Higgs unification models, results in further constraints on their parameters \cite{next}. For instance, this leads to the clear prediction of a reduced di-photon signal, due to the extended fermion sector, also in the model featuring a \textbf{10}, in analogy to the findings in the MCHM$_5$ (which had been independent of possible UV completions). The clear correlations found e.g. between the $\gamma\gamma$ and $\tau\tau$ channels, especially in the model corresponding to the MCHM$_5$, offer a nice possibility to discover or exclude the setups.
We should notice here that additional effects arising from the non-linearity of the Higgs in complete composite 
models might lead to slight modifications of the previous picture like a possible enhancement for weak-boson induced processess. 
We have studied these effects in Section \ref{sec:vf}.
Nevertheless, finding a slight reduction in the $gg\to h \to \gamma\gamma$ channel - which is still not excluded, considering the errors of current measurements - together with a depletion in the $h\tau\tau$ vertex could be interpreted as a hint for the compositeness of the $\tau$ lepton. 
Note that the phenomenology is different from the one of other extra-dimensional realizations of TeV-scale physics, like general Randall-Sundrum models \cite{Djouadi:2007fm,Azatov:2009na,Bouchart:2009vq,Casagrande:2010si,Azatov:2010pf,Goertz:2011hj,Carena:2012fk}.
As we have seen that large signals are not to be expected from the quark sector (in tentative agreement with LHC measurements) it could be the unexpected compositeness of the $\tau$ lepton that leads to first signals of compositeness  in Higgs physics at the LHC.

\paragraph{Addendum}

After completion of this work, new Higgs data were presented at the Moriond Conference \cite{:Mor}. While the significance of the excess in $pp \to h \to \gamma\gamma$ remained at the same level for the ATLAS experiment, the CMS results are now in agreement with the SM prediction within $1\,\sigma$.
Concerning the decay into $\tau$ leptons, the new ATLAS data are still consistent with a vanishing signal (with a reduced central value), whereas in CMS this decay mode has been established at the level of $3\, \sigma$. For the latter experiment the central value is essentially at the SM prediction.

If the new CMS results on $pp \to h \to \gamma\gamma$ are confirmed by ATLAS, this would lead to a better agreement with the predictions of the considered (composite) UV completions of our low energy models, see e.g. Figure \ref{fig:gamtaun}. 
However, if the $\tau \tau$ channel converges to the SM prediction, this would constrain significantly the scenarios studied in this work.

\paragraph{Acknowledgements}

We are grateful to Babis Anastasiou for stimulating discussions and Alex Azatov, Adam Falkowski, Giuliano Panico, as well as José Santiago for useful comments. The research of the authors is supported by the Swiss National Foundation under contract SNF 200021-143781. 

\appendix

\section{Form Factors}
\label{app:formfactors}

The form factors $A_{q,W}^h (\tau)$ which describe the effects of fermion 
and $W^\pm$-boson loops in the production and the decay of the Higgs boson are given by \cite{Djouadi:2005gi}
\begin{align}
    A_{q}^h (\tau) & = \frac{3 \hspace{0.25mm} \tau}{2} \left [
      \hspace{0.25mm} 1 + \left ( 1 - \tau \right ) f (\tau)
      \hspace{0.25mm} \right ] \,, \nonumber\\
    A_{W}^h (\tau) & = -\frac{3}{4} \left [ \hspace{0.25mm} 2 + 3 \tau
      + 3 \tau \left ( 2 - \tau \right ) f (\tau) \hspace{0.25mm}
    \right ]\,.
\end{align}
The function $f(\tau)$ reads
\begin{align}
    f(\tau) & = \begin{cases} - \displaystyle \frac{1}{4} \left [ \,
        \ln \left ( \displaystyle \frac{1 + \sqrt{1 - \tau}}{1 -
            \sqrt{1 - \tau}} \right ) - i \pi \, \right ]^2 \,, & \tau
      \leq 1 \,, \\[4mm] \arcsin^2 \left ( \displaystyle
        \frac{1}{\sqrt{\tau}}
      \right ) \,, & \tau > 1 \,. \end{cases}\,
\end{align}

\bibliographystyle{JHEP}
	\bibliography{myrefs}{}

\end{document}